\documentclass[aps,prl,reprint,amsmath,amssymb,superscriptaddress]{revtex4-2}
\usepackage{lineno}
\usepackage{bm}
\usepackage[usenames,dvipsnames]{xcolor}
\usepackage[bookmarks=true,colorlinks,citecolor=blue,urlcolor=blue]{hyperref}
\usepackage{graphicx}
\usepackage{mathrsfs}
\usepackage{MnSymbol}
\usepackage[normalem]{ulem}
\usepackage{times}
\usepackage{bbm}
\usepackage{relsize}
\newcommand{\correspondingauthor}{\thanks{Corresponding author: stefan.birnkammer@tum.de}}
\newcommand{\hpa}{h_{\parallel}}
\newcommand{\hpe}{h_{\perp}}

\newcommand{\dd}{{\rm d}}
\newcommand{\be}{\begin{equation}}
\newcommand{\ee}{\end{equation}}

\newcommand{\rh}{{\rm h}}

\newcommand{\abs}[1]{\vert #1 \vert}
\newcommand{\meas}[1]{\mathrm{d}#1}
\newcommand{\eqw}[1]{(\ref{#1})}
\newcommand{\eq}[1]{Eq.~(\ref{#1})}

\newcommand{\eqsto}[2]{Supplementary Eqs.~(\ref{#1}) to (\ref{#2})}

\newcommand{\fig}[1]{Fig.\thinspace{}\ref{#1}}

\newcommand{\fc}[1]{({#1})}
\newcommand{\figc}[2]{Fig.\thinspace{}\ref{#1}\thinspace{}\fc{#2}}
\newcommand{\figcc}[3]{Fig.\thinspace{}\ref{#1}\thinspace{}\fc{#2} and \fc{#3}}
\newcommand{\figcs}[3]{Fig.\thinspace{}\ref{#1}\thinspace{}\fc{#2} - \fc{#3}}

\newcommand{\iu}{\mathrm{i}}

\newcommand{\Sum}{\displaystyle\sum\limits}

\newcommand{\Prod}{\displaystyle\prod\limits}

\newcommand{\ket}[1]{| #1 \rangle}
\newcommand{\bra}[1]{\langle #1 |}

\graphicspath{{Figures/}}

\begin{document}
\newcommand{\titleinfo}{
Prethermalization in one-dimensional quantum many-body systems with confinement}

\title{\titleinfo}

\author{Stefan Birnkammer}\correspondingauthor
\affiliation{Department of Physics, Technical University of Munich, 85748 Garching, Germany}
\affiliation{Munich Center for Quantum Science and Technology (MCQST), Schellingstr. 4, D-80799 M{\"u}nchen, Germany}

\author{Alvise Bastianello}
\affiliation{Department of Physics, Technical University of Munich, 85748 Garching, Germany}
\affiliation{Munich Center for Quantum Science and Technology (MCQST), Schellingstr. 4, D-80799 M{\"u}nchen, Germany}

\author{Michael Knap}
\affiliation{Department of Physics, Technical University of Munich, 85748 Garching, Germany}
\affiliation{Munich Center for Quantum Science and Technology (MCQST), Schellingstr. 4, D-80799 M{\"u}nchen, Germany}

\begin{abstract}
Unconventional nonequilibrium phases with restricted correlation spreading and slow entanglement growth have been proposed to emerge in systems with confined excitations, calling their thermalization dynamics into question. Here, we show that in confined systems the thermalization dynamics after a quantum quench instead exhibits multiple stages with well separated time scales. As an example, we consider the confined Ising spin chain, in which domain walls in the ordered phase form bound states reminiscent of mesons. The system first relaxes towards a prethermal state, described by a  Gibbs ensemble with conserved meson number. The prethermal state arises from rare events in which mesons are created in close vicinity, leading to an avalanche of scattering events. Only at much later times a true thermal equilibrium is achieved in which the meson number conservation is violated by a mechanism akin to the Schwinger effect. The discussed prethermalization dynamics is directly relevant to generic one-dimensional, many-body systems with confined excitations.
\end{abstract}

\maketitle
Non-equilibrium states of quantum many-body systems play an important role in various fields of physics, including cosmology and condensed matter. Of particular interest is the time evolution of interacting quantum many-body systems that are well isolated from their environment~\cite{Bloch2008, Polkovnikov2011}. This research has been fueled by the progress in engineering coherent and interacting quantum many-body systems which made it possible to experimentally study unconventional relaxation dynamics. A recent interest is to explore phenomena from high-energy physics with synthetic quantum systems in a controlled way; for example lattice gauge theories have been realized~\cite{Martinez2016, Schweizer2019, Goerg2019, Yang2020a, Mil2020, Satzinger2021, Semeghini2021} and phenomena akin to quark confinement have been explored~\cite{Martinez2016, Simon2011, Tan2021, Yang2020a, Vovrosh2021b}, with great emphasis on the atypical nonequilibrium features of confined systems.
Confinement strongly affects the relaxation dynamics of the system, leading to unconventional spreading of correlations and slow entanglement growth~\cite{Kormos2017, Alvaredo2020, Scopa2021}, with striking signatures in the energy spectrum reminicent of quantum scars \cite{Robinson2019, James2019}. In spite of many efforts, a proper characterization of the full many-body dynamics and thermalization in confined systems remaines elusive so far.
\begin{figure}[t!]
\centering
	\includegraphics[width=1.0\columnwidth]{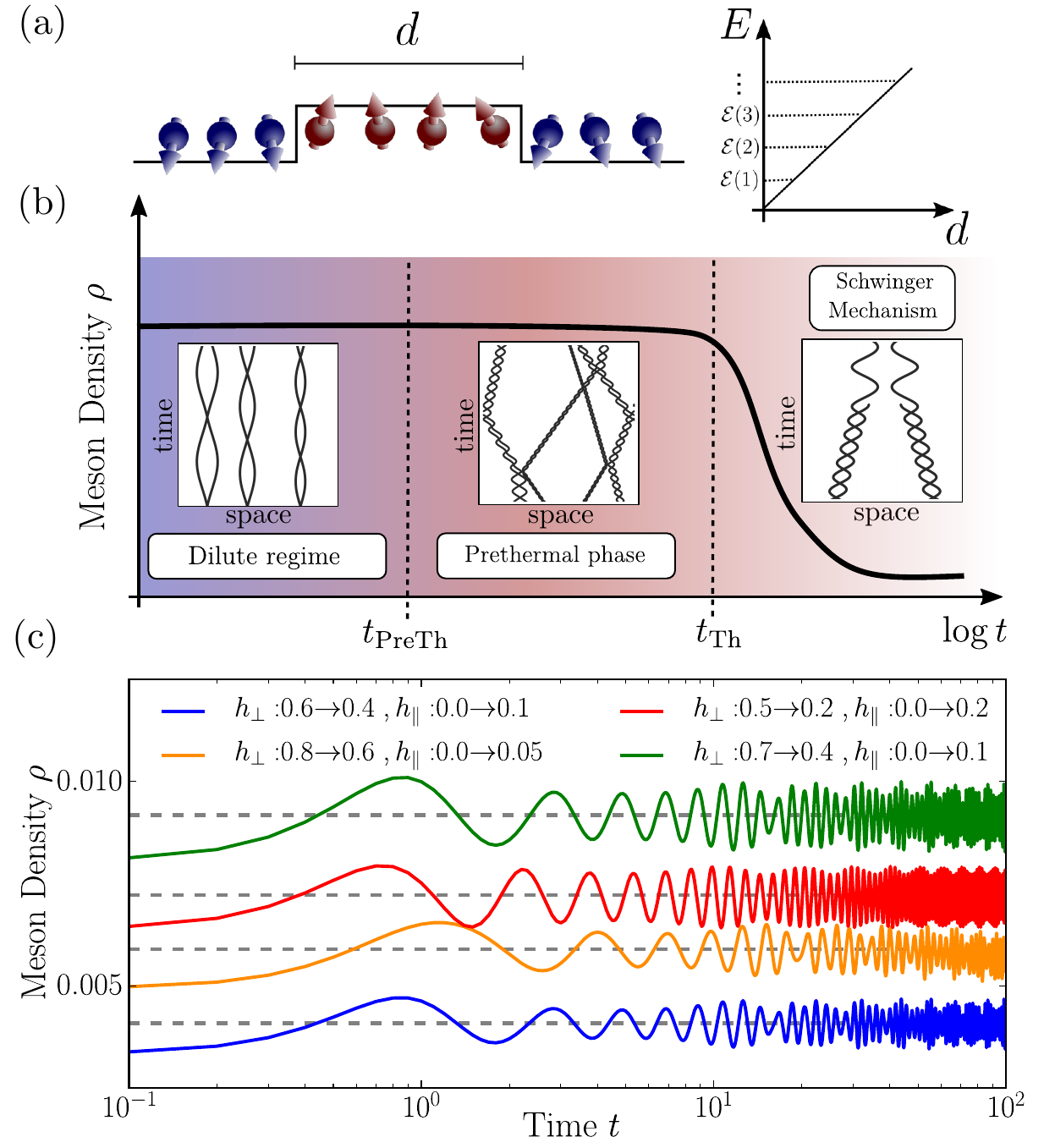}
	\caption{
 \textbf{Dynamics in the confined Ising chain.} $(a)$ Pairs of domain walls, interpreted as mesons, are confined by the longitudinal field.
    $(b)$ For weak quantum quenches of the transverse and longitudinal fields the Ising chain exhibits a multi-stage relaxation dynamics. Insets: typical domain wall trajectories in the different dynamical phases. At short times, $t<t_\text{PreTh}\propto \rho^{-2}\hpa$ (with $\rho$ the density of mesons) a metastable state arises in which mesons are at rest and well separated. For intermediate times  $t_\text{PreTh}<t<t_\text{Th}$, rare events initiate avalanches of scattering processes, leading to   prethermal Gibbs ensemble with conserved density of mesons $\rho$. At late times $t>t_\text{Th}\propto \exp[(...)\hpa^{-1}]$, the Schwinger mechanism breaks the meson number conservation leading to full thermalization.
 $(c)$ The meson density $\rho$, computed with tensor network simulations, relaxes to the analytical prediction (dashed gray lines) of Ref. \cite{Scopa2021} (see also supplementary information \cite{Note1}).}
\label{fig_mes_cons}
\end{figure}

An archetypical model to study confinement phenomena in condensed matter settings is the Ising model with both transverse and longitudinal magnetic fields~\cite{McCoy1978,Delfino1996,Rutkevich2008,Fonseca2003, Lake2010, Coldea2010}. In this model, domain walls---interpreted as quarks---are pairwise confined into mesons by a weak longitudinal field; see \figc{fig_mes_cons}{a}. 
A key feature of the model is the long lifetime of mesons, ascribed to a strong suppression of the Schwinger mechanism~\cite{Schwinger1951,Liu2019,Lagnese2021,Tortora2020,Rigobello2021,Verdel2020}, which creates new quarks from the energy stored in the confining force and viceversa.
Hence, except for some fine-tuned regimes~\cite{Karpov2020,Milsted2021,Sinha2021}, mesons are stable excitations.
Due to the approximate conservation of the meson number, various exotic dynamical phenomena have been proposed, including Wannier-Stark localization~\cite{Lerose2020, Mazza2019, Pomponio2021} and time crystals~\cite{Collura2021}.
Even though the realization of these phenomena does not require particular fine tuning, they arise in a regime in which interactions between mesons are extremely unlikely.
The few-meson scattering has been recently considered \cite{Surace2021, Karpov2020, Milsted2021, Vovrosh2021a}, but so far, apart from special limits \cite{Bastianello2021b}, the full many-body dynamics of confined systems has not been addressed.
Irrespective of these exciting effects, the Ising model with longitudinal and transverse fields is non-integrable~\cite{Delfino1996} and features a Wigner-Dyson level statistics of the eigenenergies~\cite{Kim2013}. Hence, one would expect on general grounds~\cite{Deutsch1991, Srednicki1994, Rigol2008} that the system thermalizes at late times and interactions between mesons can become relevant.
Given this wealth of unconventional non-equilibrium phenomena and the discrepancy with the expected thermalization in non-integrable models, it is important to understand the mechanisms of relaxation and their timescales.

In this work, we investigate the relaxation dynamics of one-dimensional systems in the presence of confinement, with focus on the Ising chain as a primary example.
Two scenarios could be envisioned for the  thermalization process. 
The first one is that
the Schwinger effect, leading to a violation of the meson number conservation, could be the only responsible mechanism for equilibration, causing an extremely slow thermalization dynamics.
A more exciting, second scenario involves an intermediate thermalization of the mesons themselves. Here, we show that indeed the second scenario is realized. Generic states first relax to a Gibbs ensemble in which the meson number is conserved up to extremely long times; \figcc{fig_mes_cons}{b}{c}.
We show that relaxation to this state is activated through rare events in which two mesons are produced in their vicinity, initiating an avalanche of scattering events. This prethermal state can then be understood as a dilute thermal gas of mesons with conserved meson density. Only at exponentially long times, the Schwinger mechanism causes a full thermalization of the system coupling sectors with a different number of mesons. While we choose to focus on the Ising chain as the simplest example where both analytical and numerical progress can be made efficiently, our findings can be extended to generic confined many-body systems as we emphasize in the discussion section.
\ \\ \ \\
{\large\textbf{Results}}\\
\textbf{Model and protocol.} The Ising chain with both transverse and longitudinal fields is described by the Hamiltonian
\be\label{eq_H_tilted}
\hat{H}=-\sum_j[ \hat{\sigma}^z_{j+1}\hat{\sigma}_j^z+\hpe \hat{\sigma}_j^x +\hpa \hat{\sigma}_j^z]\, .
\ee
In the pure transverse-field regime ($\hpa=0$) the model is equivalent to non-interacting fermions and exhibits spontaneous $\mathbb{Z}_2-$symmetry breaking for $|\hpe|\le 1$ in the thermodynamic limit. 
For $\hpe \to 0$, the two degenerate ground states $|\text{GS}_\pm\rangle$ are simple product states of maximally positive/negative magnetization, which are renormalized for finite transverse field, such that $\langle \text{GS}_\pm|\hat{\sigma}_j^z|\text{GS}_\pm\rangle=\pm \bar{\sigma}$, with $\bar{\sigma}=(1-\hpe^2)^{1/8}$ \cite{Sachdev2011}.
In this phase, the fermionic modes are interpreted as (dressed) domain walls (or kinks) relating the two vacua and are thus of topological nature.
A small longitudinal field $\hpa>0$ lifts the ground state degeneracy, leading to a low-energy ``true vacuum" and a high-energy ``false vacuum," and induces a pairwise linear potential $\propto 2 \hpa \bar{\sigma}$ between kinks; \figc{fig_mes_cons}{a}.
\begin{figure}[t!]
    \includegraphics[width=1.01\columnwidth]{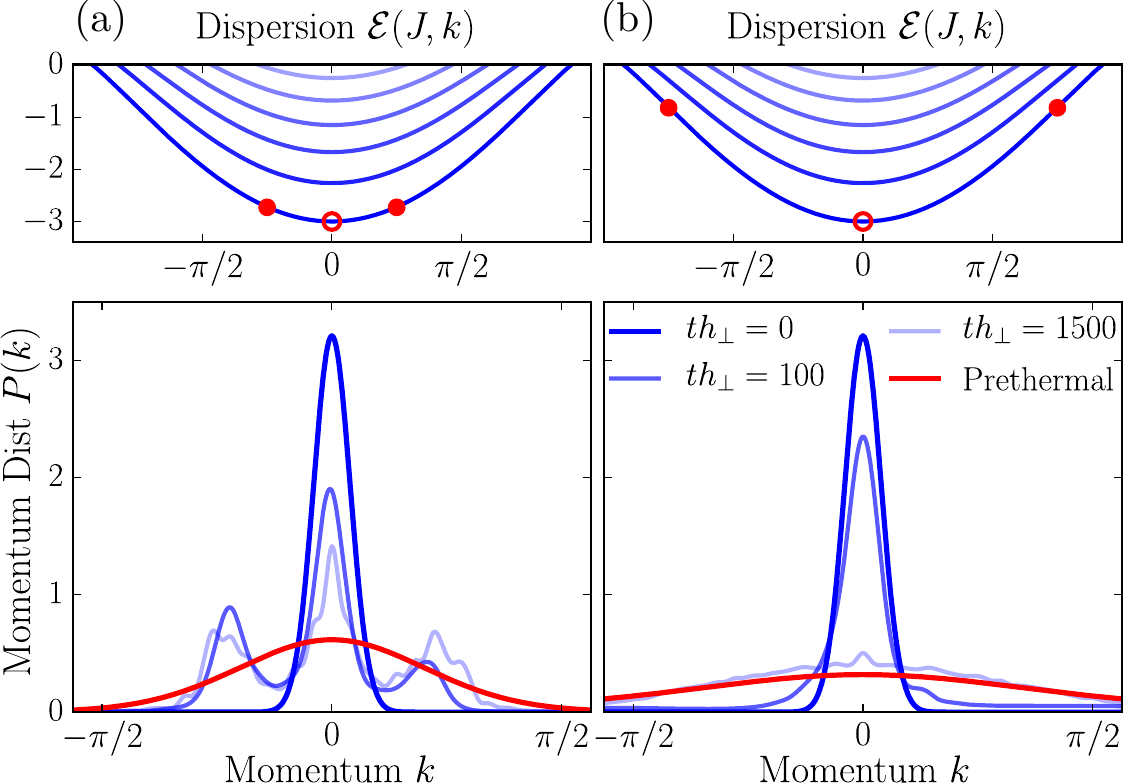}
\caption{
	\textbf{Prethermalization of quantum mesons.}
	We create three mesons with Gaussian wave packets tuned to target the lowest energy band at momenta $\{-k_0,0,k_0\}$ in a chain of length $L=100$ with confinement field $\hpa/\hpe=0.1$ (see Methods). Energy bands are computed through a numerical solution of the two-fermions quantum Hamiltonian \cite{Note1}.
	The evolution of the momentum distribution $P(k)$ of the meson initialized at rest (empty circle in upper panels) is shown.
$(a)$ For $k_{0}=\pi/4$ the energy of the initial wave packets is below the second band of the single meson dispersion (upper panel), which causes meson scattering events to be elastic and prohibits relaxation to a prethermal ensemble (red curve). $(b)$ For $k_{0}=3\pi/4$, the initial wave packets are resonant with other bands, which can then be populated, and lead to inelastic scattering. This quantum state relaxes to a prethermal configuration at late times.
Prethermal curves are computed according to the hard-rods thermodynamics \eqref{eq:thermoclass}  using the exact quantum eigenfunctions rather than the semiclassical prediction \cite{Note1}.
}
	\label{fig_ED}
\end{figure}

We consider the following quantum quench~\cite{Kormos2017}: The system is initialized for $\hpa=0$ in one of the two degenerate ground states (specifically, we select $\langle \hat{\sigma}^z\rangle>0$) and then brought out of equilibrium by suddenly changing both the transverse and the longitudinal field components.
Building on the knowledge of quenches in the transverse field only \cite{Calabrese2011}, one can argue that fermions are locally produced in pairs with opposite momenta \cite{Kormos2017}, each of them having a dispersion $\epsilon(k)=2\sqrt{(\cos k-\hpe)^2+\sin^2 k}$. However, pairs of fermions are then confined due to the finite longitudinal field $\hpa\ne 0$. 
For weak quenches, very few excitations are produced and, due to translational invariance, mesons are mostly initialized at rest and are well-isolated. Their stability is guaranteed by the strong suppression of fermion number-changing processes.
In the case of small transverse field ($|\hpe|<1/3$) two fermions cannot energetically couple to the four-fermion sector without using the energy stored in the false-vacuum string.
Hence, this process resembles the false vacuum decay, whose lifetime has been shown to scale exponentially with $\hpa^{-1}$~\cite{Schwinger1951}.
Even in the less restricted regime where the scattering of two fermions into four is energetically allowed $(1/3<|\hpe|<1)$, the cross section is induced by the weak longitudinal term, leading to a meson lifetime that scales algebraically in the longitudinal field $\hpa^{-3}$~\cite{Rutkevich2005}. To confirm this expectation, we perform tensor network simulations \cite{Murg2008, Schollwoeck2011, Hauschild2018} based on the TenPy library~\cite{Hauschild2018} of the quantum quench and compute the meson density $\rho$; \figc{fig_mes_cons}{c}. We checked convergence of our data with bond dimension on the shown time scales (data is shown for $\chi = 256$). In the limit of small $\hpa$ the meson number is conserved on the numerically accessible time scales (see Methods).
\ \\ \ \\
\textbf{Excitation spectrum and thermodynamics.} Assuming that the meson number  is conserved, we now study the thermodynamics of a gas of mesons, which is expected to describe the prethermal state.
In the dilute regime, the mean-free path is much larger than the typical meson length. In a first approximation, we therefore neglect the effects that the size of the meson has on the thermodynamics.
A convenient starting point is the semiclassical limit of a single meson, in which one treats the two fermions as point-like particles with coordinates $(x_{1,2},k_{1,2})$ governed by the classical Hamiltonian
\be\label{eq_cl_ham}
\mathcal{H}=\epsilon(k_1)+\epsilon(k_2)+2\hpa\bar{\sigma}|x_1-x_2|\, .
\ee 
The semiclassical approximation holds when interactions cannot resolve the discreteness of the underlying lattice, i.e., for $\hpa\ll 1$. Hence, the position of the particle $x_{1,2}$ is a continuous variable.
In the reduced two-body problem, the total momentum $k=k_1+k_2$ of a meson is conserved, thus the dynamics of the relative coordinates $(q=(k_1-k_2)/2,x=x_1-x_2)$ is governed by $\mathcal{H}_{\text{rel}}(q,x)=\epsilon(k/2+q)+\epsilon(k/2-q)+2\bar{\sigma} \hpa |x|$. 
Then, the thermal probability of having a meson with a given energy and momentum is $P(E,k)=e^{-\beta (E-\mu)}\int \frac{\dd q\dd x}{(2\pi)^2} \,\delta(\mathcal{H}_\text{rel}(q,x)-E)$, where the inverse temperature $\beta$ and chemical potential $\mu$ must be fixed by matching the initial average energy and meson density, respectively.
The integral over the relative coordinates is most conveniently tackled by transforming to action-angle variables $(J,\phi)$ \cite{Goldstein2011}, where $J\equiv \oint_{\mathcal{H}_{\text{rel}}(q,x)=\mathcal{E}(J,k)} q\dd x $ labels the phase-space orbits of the classical motion and $\phi$ is a periodic variable $\phi\in[0,1]$, leading to $P(E,k)=e^{-\beta (E-\mu)}\int\frac{\dd J}{(2\pi)^2} \delta(\mathcal{E}(J,k)-E)$.
Leaving the classical limit, the energy levels become quantized according to the Bohr-Sommerfeld rule  $J=2\pi(n-1/2)$, where $n$ is a natural number \cite{Rutkevich2008}.

Away from the dilute regime mesons should be treated as extended objects and their thermodynamics needs to be suitably modified.
To this end, we consider mesons as hard-rods of fixed length $\ell(J,k)$, the latter being the meson length averaged over one oscillation period.
Within this assumption, $P(E,k)$ gets modified as
\be
\label{eq:thermoclass}
\frac{P(E,k)}{1-\rho M}=e^{-\beta(E-\mu)}\int\frac{\dd J}{(2\pi)^2} \delta(\mathcal{E}(J,k)-E)e^{-\rho \ell(J,k)(1-\rho M)^{-1}}\,
\ee
with $\rho$ the meson density and $M$ the average meson length, which are self-consistently determined by $P(E,k)$;
see also supplementary information \footnote{Supplementary information for details on the confining dynamics; characterization of the prethermal state; initialization of moving mesons by staggered field pulses; further information on details of numerical simulations.}. 
The meson coverage $\rho M$ is connected to the magnetization of the Ising chain as $\rho M=1/2- \bar{\sigma}^{-1}\langle S^z\rangle$.

While we chose to present the thermodynamics from the semiclassical perspective for the sake of clarity, quantum effects can be important when the fermion bandwidth becomes comparable with the longitudinal field and the Born-Sommerfeld quantization is a poor approximation. In this regime, the classical Hamiltonian \eqref{eq_cl_ham} can be directly promoted to a quantum object and explicitly diagonalized \cite{Scopa2021}, thus replacing the $J-$integration in Eq. \eqref{eq:thermoclass} with a discrete sum \cite{Note1}.  
\ \\ \ \\
\textbf{Prethermalization of quantum mesons.} In order to show that meson-meson scattering leads to a prethermal Gibbs Ensemble, we numerically calculate the  time evolution in the subspace with a fixed number of mesons using exact diagonalization (see Methods). We consider a chain of length $L$ with periodic boundary conditions, and focus on the limit $0<\hpe\ll 1$ where fermions can be identified with domain walls.
In this regime, $\bar{\sigma} \to 1$ and the confinement strength is determined by $\hpa/\hpe$.
We initialize the state in the form of moving wave packets and probe relaxation by tracking the meson momentum distribution; \fig{fig_ED}. Whereas for two mesons, energy and momentum conservation inhibits thermalization, see supplementary information \cite{Note1}, for three mesons we observe the relaxation to the prethermal Gibbs ensemble; \eq{eq:thermoclass}. Two-body scattering processes between different energy bands are responsible for the thermalization; \figc{fig_ED}{b}. 
For wave packets which are initialized with energies below the second band thermalization is largely suppressed, as two-body collisions become elastic due to momentum-energy conservation and three-body scattering events are unlikely; \figc{fig_ED}{a}. We provide additional details on the thermalization in the supplementary information \cite{Note1}.
\begin{figure}[t!]
	\includegraphics[width=1.01\columnwidth]{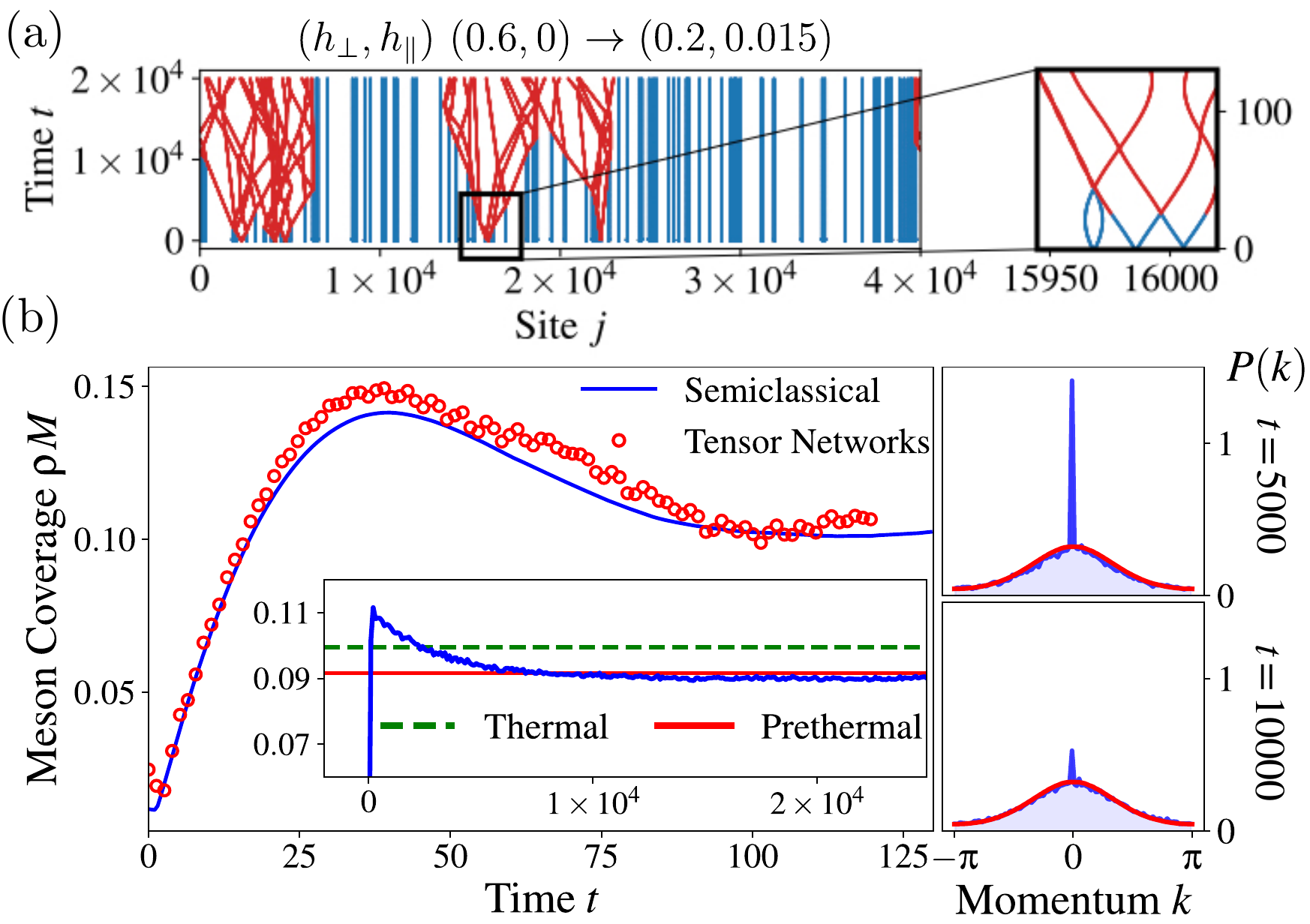}
	\caption{
	\textbf{Prethermalization through rare events.}
	$(a)$ Typical semiclassical trajectories obtained with the Truncated Wigner Approximation. Most fermions belong to mesons at rest (blue lines), but rare events in which mesons are in close vicinity lead to an avalanche effect putting mesons in motion (red lines) and activating dynamics in the entire meson ensemble. $(b)$ For comparably small values of $\hpa$ semiclassical results for the average meson coverage $\rho M$ agree well with exact quantum evolution obtained from tensor network techniques. (Inset) The semiclassical analysis reveals relaxation towards a prethermal plateau (red), which is distinct from the thermal state in the absence of meson conservation (green dashed). Side panels: Relaxation of the semiclassical ensemble is also reflected in the decay of the the momentum distribution $P(k)$ at $k\approx 0$. Thermal ($\mu=0$) and prethermal ($\mu$ fixed by number of mesons) predictions are computed with Eq. \eqref{eq:thermoclass}, directly in the classical limit.}
	\label{fig_semiclassic_1}
\end{figure}
\ \\ \ \\
\textbf{Prethermalization through rare events.} Equipped with the meson conservation, the thermodynamics of the prethermal state, and the quantum thermalization of a few mesons, we now study the full quench protocol. In order to access large system sizes and time scales, we use the Truncated Wigner Approximation \cite{Polkovnikov2010} on the quantum dynamics projected in the fermion number conserving sector (see Methods).
In order to study the relaxation dynamics, a precise knowledge of the excitation content of the initial state is crucial. The quantum quench of both the longitudinal and the transverse field excites dilute pairs of fermions with opposite momenta $(k,-k)$ at density $n(k)$, which can be computed from the quench parameters~\cite{Kormos2017,Scopa2021}. These pairs of fermions are then confined into mesons by the longitudinal field according to Eq. \eqref{eq_cl_ham}.

For small quenches within the ferromagnetic phase, the  density of mesons is low. Typically, mesons are excited far apart and are thus isolated and at rest. In this scenario, inter-meson scattering and thermalization seems impossible. However, considering only the typical behavior is misleading, as the probability of creating two nearby mesons is never strictly zero. To obtain a rough estimate, we consider the maximum size $d_\text{max}$ that a meson can have when fermions are initially created at the same position, which is given by $d_\text{max}=4\hpe/(\hpa \bar{\sigma})$, and compare it with the meson density $\rho$.
On a finite volume $L$, the probability $P(L)$ that $N=L\rho$ randomly distributed particles are placed at distance larger than $d_\text{max}$ is $P(L)=\frac{1}{L^N}\prod_{j=0}^{N-1}(L-j d_\text{max})\simeq e^{-L\rho^2 d_\text{max}/2}$. No matter how small the excitation density $\rho$ is, eventually in the thermodynamic limit the probability that all the excited mesons are far apart vanishes.
Crucially, the rare near-by mesons scatter and acquire a finite velocity. These moving mesons consequently trigger an avalanche, that hits the surrounding mesons, and initiates prethermalization; see \figc{fig_semiclassic_1}{a} for a typical meson configuration.
In \figc{fig_semiclassic_1}{b}, left, we first ensure that the semiclassical approximation is reliable for the chosen quench parameters, by comparing with tensor network simulations on the reachable time scale (convergence with bond dimension is checked; data shown for $\chi = 256$).
Then, we use the semiclassical approach to probe extremely long times, observing prethermalization of both the meson coverage $\rho M$ (inset) and the momentum distribution of mesons $P(k)$ (right panels). For the latter the initial $\propto\delta(k)$ peak decays due to the aforementioned avalanche effect and relaxes to a smooth prethermal Gibbs distribution.

The density dependence of the prethermalization timescale $t_\text{PreTh}$ can be understood as follows.
Initially, the configuration consists of large regions of average size $\sim (\rho^2 d_\text{max})^{-1}$ with mesons at rest separated by growing thermalizing domains. Hence, we estimate prethermalizing regions to cover the whole system on a typical time $t^* \sim (\rho^2 d_\text{max} {\rm v})^{-1}$, where ${\rm v}$ is a typical velocity. Once all mesons are set in motion, two-body inelastic scatterings drive the relaxation of the system on a timescale $t^{**}\sim (\rho {\rm v})^{-1}$. At low excitation density, $t^*\gg t^{**}$, hence $t_\text{PreTh}\sim (\rho^2 d_\text{max}{\rm v})^{-1}$. 
\begin{figure}[t!]
	\includegraphics[width=1.01\columnwidth]{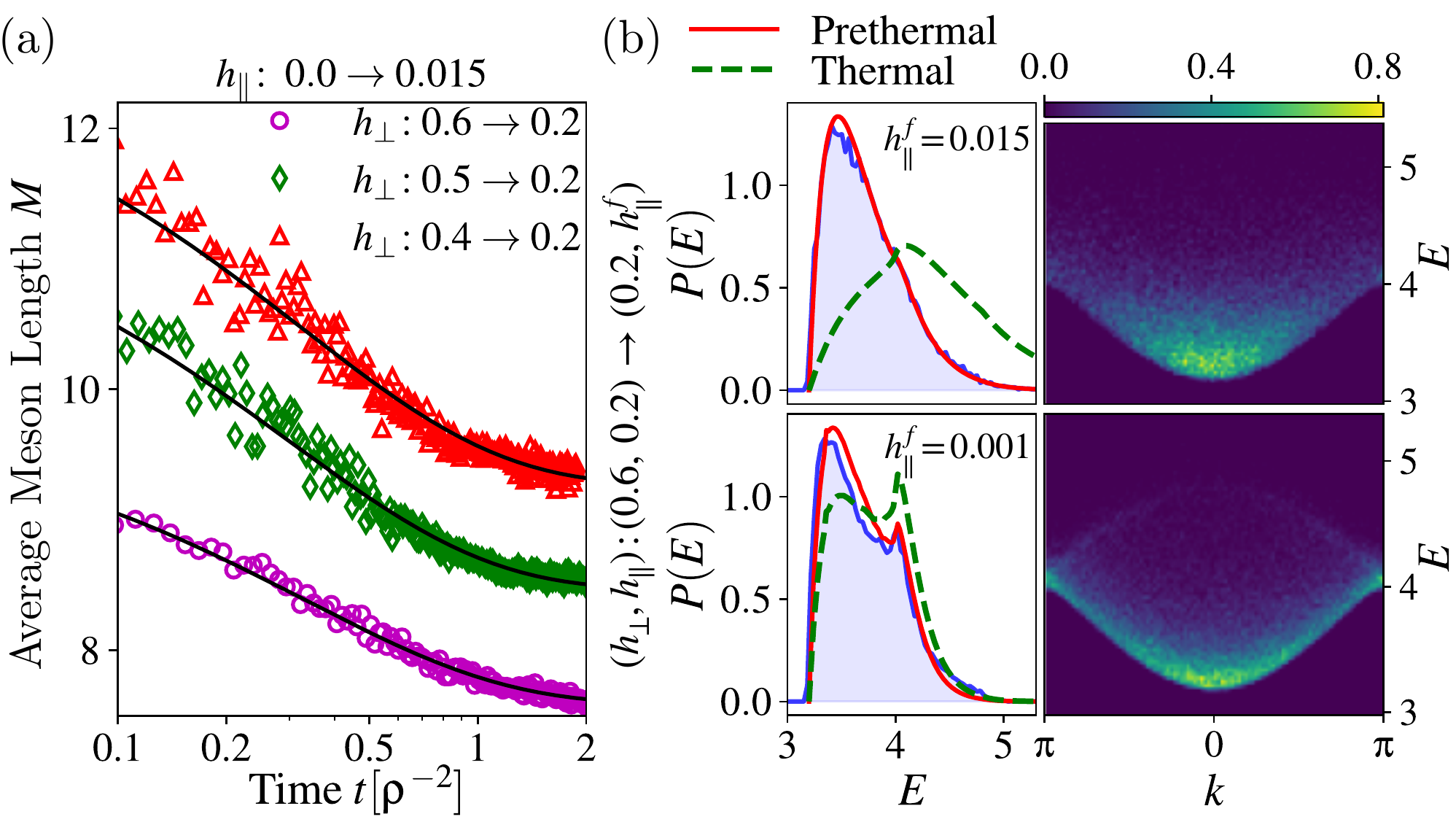}
	\caption{\textbf{Approaching the prethermal state and thermodynamics.}
	$(a)$ The late time relaxation of the average meson length $M$ to the prethermal plateau is well-described by the  prediction $\langle \mathcal{O}(t)\rangle=\mathcal{O}_\text{PreTh}+\Delta \mathcal{O} F(t{\rm v}d\rho^2 )$. The quantity ${\rm v}d$ is obtained from a fit to the data.
	$(b)$ The normalized one-meson phase space occupation relaxes to a prethermal ensemble (prethermal: red continuous line; thermal: green dashed line; numerics: blue shaded area).
	Finite-density corrections are captured by the hard-rods approximation and cause an additional peak in the energy distribution $P(E)$ (bottom).
	The relative difference in the meson densities between the thermal and prethermal ensemble $\Delta \rho = (\rho_\text{PreTh}-\rho_\text{Th})/\rho_\text{Th}$ are $\Delta\rho=0.16$ and $\Delta \rho=\mathcal{O}(10^{-3})$ for $\hpa^f=0.015$ (top) and $\hpa^f=0.001$ (bottom), respectively. Thermal and prethermal observables are computed with Eq. \eqref{eq:thermoclass}, real-time evolution is obtained within the Truncated Wigner Approximation.
	}
	\label{fig_semiclassic_2}
\end{figure}

Building on this approximation, we can understand the relaxation of local observables by assuming that prethermalizing regions contribute with $\mathcal{O}_\text{PreTh}$, while regions with static mesons retain the initial value $\mathcal{O}_0$ (after a short dephasing time). Hence, $\langle \mathcal{O}(t)\rangle$ follows the average growth of thermalizing regions
\be\label{eq_S_ext2}
\langle \mathcal{O}(t)\rangle=\mathcal{O}_\text{PreTh}+\Delta \mathcal{O}\int_0^\infty \dd D \, \mathcal{P}(D) \frac{D- 2{\rm v} t}{D}\theta(D-2{\rm v} t)\, ,
\ee
where we approximate each prethermalizing region to growth in a lightcone fashion with velocity ${\rm v}$ and $\Delta \mathcal{O}=\mathcal{O}_0-\mathcal{O}_\text{PreTh}$
Above, $\theta(x)$ is the Heaviside theta function $\theta(x>0)=1$ and zero otherwise, $D$ is the distance between two rare events which is distributed with probability distribution $\mathcal{P}(D)$.
By its very definition, $\mathcal{O}_0$ can be computed as the late-time limit of the single meson approximation, since outside of the scrambling region the mesons are not interacting. Finally, the term $\frac{D- 2{\rm v} t}{D}$ is nothing else than the portion of frozen region 
that remained after the thermalizing region propagated with velocity ${\rm v}$ inside of it. 
The last step is now to estimate $\mathcal{P}$. We have already computed the probability that, within a system of size $L$, there are no rare events. In the computation, we used the maximum extension of a meson $d_\text{max}$ as an upper bound, but a better estimate is obtained using the average size of the excited mesons, which we call $d$. Hence, the probability that within an interval $L$ there are no rare events is $P(L)= e^{-L \rho^2 d/2}$. The distributions $P(L)$ and $\mathcal{P}(D)$ are  related by $P(L)=\int_{L}^\infty\dd D\, \mathcal{P}(D)$, leading to $\mathcal{P}(D)=\frac{\rho^2 d}{2} e^{-D \rho^2 d/2}$. 
With this approximation, Eq. \eqref{eq_S_ext2} can be recast in a scaling form $\langle \mathcal{O}(t)\rangle=\mathcal{O}_\text{PreTh}+\Delta \mathcal{O}\, F(t{\rm v}\rho^2 d)$ with $ F(\tau)=\int_\tau^\infty \dd s\, e^{-s}(1-\tau/s)$.

The full numerical results agree with this picture; \figc{fig_semiclassic_2}{a}.
Since $d_\text{max}\propto \hpa^{-1}$, smaller longitudinal fields leads to a shorter prethermalization time scale for the same meson density $\rho$.
Even in the less favorable case where $\hpe$ is kept constant and only $\hpa$ is quenched (i.e., only the small longitudinal field is ultimately responsible of creating fermionic excitations), we find $\rho^2\propto \hpa^{-2}$ \cite{Note1}. Hence, there is in any case a separation of scales between the prethermalization time $t_\text{PreTh}\propto \hpa^{-1}$ and the violation of meson-number conservation $t_\text{Th} \sim \exp[(...)\hpa^{-1}]$, consistently ensuring the existence of the prethermal regime for a large class of quenches.

In \figc{fig_semiclassic_2}{b} we study the semiclassical prethermal regime for different confining strengths, but the same average density and energy. For $\hpa^f = 0.015$ (top) the average meson length is shorter than for $\hpa^f = 0.001$ (bottom). We observe that the larger size of the mesons influences the phase space distribution as follows: \emph{i)} it introduces a momentum-dependent cutoff in the energy, which is ultimately caused by the fact that the average meson length is bounded by the mean free path, and \emph{ii)} the probability distribution is squeezed to the boundaries of the allowed phase space. A consequence of this is the emergence of a peak in the energy distribution corresponding to the Brillouin zone boundaries (compare bottom and top distibution functions). This effect is captured by our hard-rods approximation.
For this choice of parameters, we observe the thermal number of mesons is lower than the prethermal one, hence thermalization is achieved by fusing small mesons into larger ones, i.e., by the reverse process of the Schwinger effect; \figc{fig_mes_cons}{b}.
In order to conserve the total mean energy, the thermal distribution has more high-energy mesons excited than the prethermal case.
The difference between the prethermal and thermal state is reduced at higher meson densities, where the hard-rods correction penalizes large mesons.

By virtue of the simple underlying kinetic mechanism, the validity of our study is expected beyond the classical realm to hold in the quantum case as well, with an additional refinement. As previously mentioned, thermalization is activated by two-body scattering between different energy bands. Hence, the estimate of $t_\text{PreTh}$ should be corrected considering that only a fraction of $\rho$ is contributing to the inelastic scattering.

\ \\ \ \\
{\large\textbf{Discussion}}\\
Confined spin chains exhibit an intriguing multi-stage thermalization dynamics. We show that not the Schwinger mechanism is responsible for activating transport, but rather rare events in which two mesons are generated in their vicinity lead to a prethermal regime, that can be understood as a thermal gas of mesons. The different mechanism ensures the separation of timescales and the existence of a prethermal regime.
The prethermalization time can be greatly reduced by considering quench protocols that create mesons with non-zero velocity. This, for example, can be realized with spatially-modulated pulses of the transverse field~\cite{Bastianello2018, Note1}.

We used the Ising chain \eqref{eq_H_tilted} as a prototypical model to demonstrate the rich relaxation dynamics. However, similar dynamics is expected in other confined many-body systems as well; for example lattice gauge theories~\cite{Zohar2015,Dalmonte2016,Preskill2018}. Incidentally, we notice that the Ising chain \eqref{eq_H_tilted} can be interpreted as a $\mathbb{Z}_2-$gauge theory in the zero charge sector, where matter degrees of freedom have been integrated out by virtue of the Gauss law \cite{Lerose2020}.
A prominent example of a different lattice gauge theory is the $U(1)$ quantum link model
\begin{multline}
\label{main_QLM}
    H_{\mathrm{QLM}}= -\omega \sum_{j=1}^{L-1}(\phi_{j}^{\dagger} S_{j, j+1}^{+}\phi_{j+1}^{\phantom{\dagger}} + \mathrm{h.c.}) +\\
    \frac{m}{2} \sum_{j=1}^{L} (-1)^{j} \phi_{j}^{\dagger} \phi_{j}^{\phantom{\dagger}} - 2 h_{\parallel} \sum_{j=1}^{L-1} S_{j, j+1}^{z},
\end{multline}
where staggered Kogut-Susskind fermionic matter $\phi_{j}$~\cite{Kogut1975} interacts via the gauge degrees of freedom encoded in the spin variables $S_{j,j+1}^\alpha$. In this model, (anti-)quarks correspond to defects in the staggered matter degrees of freedom and quark-antiquark pairs experience a linear confinement potential $\propto h_{\parallel}$.
In a recent work~\cite{Surace2020}, it has been understood that the Hamiltonian \eqref{main_QLM} maps to the Fendley-Sengupta-Sachdev Hamiltonian \cite{Fendley2004} describing one-dimensional Rydberg atom arrays \cite{Bernien2017} which may experimentally probe our findings. Within this implementation, the vacuum of the gauge theory is mapped into a chain where atoms are excited in their Rydberg state on even sites, then quark-antiquark pairs are excited by placing defects in this configuration. Realizing a quantum quench akin to the one studied here, will thus lead to the same multistage thermalization dynamics. Further details can be found in supplementary information \cite{Note1}.

Other experimentally relevant models with confinement can be realized in spin ladders \cite{Lagnese20,Lagnese22,Rutk22} or long-range systems \cite{Liu2019,Tan2021}.
Particularly intriguing features can be expected for long-range models: in contrast to the short-range Ising chain \eqref{eq_H_tilted} and the quantum link model \eqref{main_QLM} discussed above, long-range couplings induce slowly-decaying (power-law) interactions between mesons which cannot be neglected.
The long-range interactions can be envisaged to affect the approximation of dilute mesons, rendering prethermalization faster on the one hand, but making the approximation of the prethermal regime as a thermal gas of non-interacting mesons unreliable on the other hand. It would be interesting to extend our prethermal description to capture meson-meson interactions.

Another intriguing direction would be to address scenarios where the violation of the meson number conservation is not negligible and must be properly considered. Can one observe and describe the drift to the thermal regime in such cases? A kinetic theory would require a quantitative understanding of meson creation-annihilation processes beyond the estimates discussed in this work.

\ \\ \ \\
{\large\textbf{Methods}}\\
\textbf{Tensor network simulations.} 
We used tensor network simulations to demonstrate the conservation of the meson number during the quantum evolution. Whereas time evolution can be carried out using the standard method of infinite Time-Evolving Block Decimation (iTEBD)~\cite{Schollwoeck2011, Hauschild2018}, measurements of the meson number are more subtle. We outline how the mesonic number operator can be embedded efficiently in tensor network formalism.\\
The construction relies on the exact solution of the transverse Ising model, which we summarize in supplementary information \cite{Note1}.
Let  $\{\gamma_{k}, \gamma_{k}^{\dagger}\}$ be the fermionic creation and annihilation operators that diagonalize the transverse field Ising model in the absence of a longitudinal field $\hpa=0$, the mesonic number operator is obtained as half of the mode number operator

\begin{align}
\label{eq:MesonNumberOperator}
\nonumber 2N_\text{mes} &=\int_{-\pi}^{\pi}\frac{\meas{k}}{2\pi} \gamma_{k}^{\dagger}\gamma_{k}^{} = \int_{-\pi}^{\pi}\frac{\meas{k}}{2\pi} \big(\cos^{2}{\theta_{k}} - \sin^{2}{\theta_{k}} \big) \alpha_{k}^{\dagger}\alpha_{k}^{} +\\
&+\delta(0)\sin^{2}{\theta_{k}} +\iu \sin{\theta_{k}}\cos{\theta_{k}} \big(\alpha_{-k}^{}\alpha_{k}^{} - \alpha_{k}^{\dagger}\alpha_{-k}^{\dagger}\big)\,,
\end{align}
where the $\hat{\alpha}_k=\cos\theta_k \hat{\gamma}_{k}+i\sin\theta_k \hat{\gamma}^\dagger_{-k}$ are the Jordan-Wigner fermions in the Fourier basis $\hat{c}_j=\int_{-\pi}^{\pi}\frac{\dd k}{\sqrt{2\pi}} e^{ik j}\hat{\alpha}_k$, which are eventually related to the original spin variables as $(\hat\sigma_j^x+i\hat\sigma_j^y)/2=\exp\big(i\pi \sum_{i<j} \hat{c}^\dagger_{i}\hat{c}_{i}\big)\hat{c}^\dagger_j$.
The Bogoliubov angle is tuned in such a way $\theta_k=-\frac{1}{2i}\log\left(\frac{h_\perp-e^{i k}}{(\cos k-h_\perp)^2+\sin^2k}\right)$.

The divergent factor $\delta(0)$ arises from equal-momentum commutation relation and it must be regularized $\delta(0) = L$, with the system size $L$.
Moving to the coordinate space, the meson number can be thus written as

\begin{multline}
\label{eq:Meson-Number-Op final}
        N_\text{mes} = \frac{1}{2} \Sum_{j} \Big\{\Sum_{\ell} \big[f_{1}(\ell)c_{j+\ell}^{\dagger}c_{j}^{} + \frac{1}{2}f_{2}(\ell)\big(c_{j+\ell}^{}c_{j}^{}+ c_{j}^{\dagger}c_{j+\ell}^{\dagger} \big)\big] \\
+ \int_{-\pi}^{\pi}\frac{\meas{k}}{2\pi}\sin^{2}{\theta_{k}}\Big\}. 
\end{multline}
In \eq{eq:Meson-Number-Op final} we introduced the functions $f_{1}(\ell), f_{2}(\ell)$ encoding the non-locality of the Jordan-Wigner mapping

\begin{align}
f_{1}(\ell)&=\int_{-\pi}^{\pi}\frac{\meas{k}}{2\pi} e^{\iu k\ell} \;\cos{2\theta_{k}}\, ,   &f_{2}(\ell)&=\int_{-\pi}^{\pi}\frac{\meas{k}}{2\pi} e^{\iu k\ell}\; \iu\sin{2\theta_{k}}.
\end{align}
Finally, we can also invert the Jordan-Wigner mapping to obtain the expression in the spin basis
$c_{j+\ell}^{\dagger}c_{j}^{} = \sigma_{j +\ell}^{-} \Big(\Prod_{i=j}^{j +\ell - 1} \sigma_{i}^{z}\Big) \sigma_{j}^{+} $ and $ c_{j+\ell}^{}c_{j}^{} = \sigma_{j +\ell}^{+} \Big(\Prod_{i=j}^{j +\ell - 1} \sigma_{i}^{z}\Big) \sigma_{j}^{+}$.

Since $N_\text{mes}$ contains in general long-ranged terms an efficient representation in terms of an MPO strongly depends on the functional form of $f_{1}(\ell), f_{2}(\ell)$. For small values of the transverse field $\hpe$ we find that both $f_{1}(\ell)$ and $f_{2}(\ell)$ can be approximated by an exponential decay for $l>0$. This enables us to make use of the efficient representation of MPOs with coefficients exponentially decaying with distance discussed e.g. in Ref.~\cite{Schollwoeck2011}. 
  \begin{figure}[t!]
    \centering
    \includegraphics[width=1.0\columnwidth]{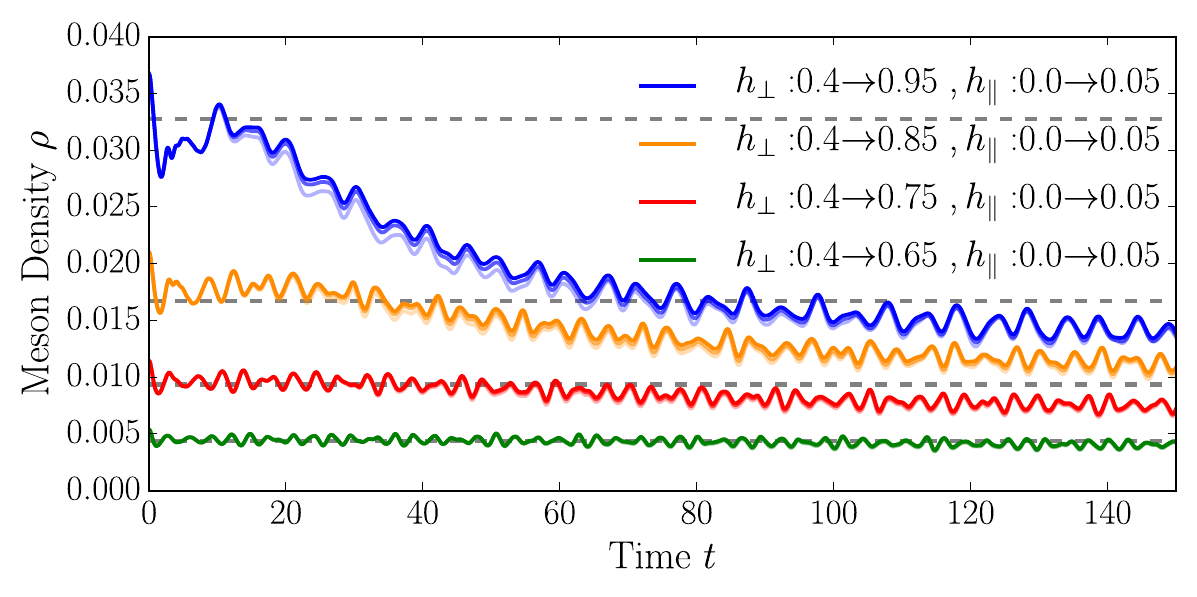}
    \caption{\textbf{Breakdown of meson number conservation.}
    We show results for a quantum quench of the ground state with initial field configuration $(h_{\perp}, h_{\parallel}) = (0.4, 0.0)$ to values of the transverse field $h_{\perp}\in\{0.65, 0.75, 0.85, 0.95 \}$ and additional confining longitudinal field $h_{\parallel}=0.05$. For all quenches we show results obtained using iTEBD time evolution for a unit cell of $L=40$ sites with bond dimensions of $\chi \in\{256, 384, 512\}$ (light to dark solid lines). We find that quenches close to the critical value of the transverse field only for a short time show the expected value of the meson number (gray dashed lines) before showing a decay in the number of mesons. The timescale, on which such a decay takes place increases and finally exceeds the numerically accessible times for quenches deep into the ferromagnetic phase $(h_{\perp}\ll 1)$.}
    \label{fig:BreakdownNumberConservation}
\end{figure}

With this method, we can analyze quantum quenches in the Ising chain and follow the evolution of the number of mesons, checking whether it is approximately well-conserved or corrections are important.
In \figc{fig_mes_cons}{c} of the main text we focus on parameter regimes where the meson number is oscillating around a constant value, in very good agreement with the analytic prediction of Ref. \cite{Scopa2021}.
Oscillations have a technical origin and are due to the fact that, strictly speaking, it is not the number of fermions that is conserved, but rather the fermion number after a perturbatively-small basis rotation \cite{Lerose2020}. Hence, in the original basis the fermion number couples to non-conserved quantities as well, which cause the small superimposed oscillations.
To complement the analysis of \figc{fig_mes_cons}{c} , in \fig{fig:BreakdownNumberConservation} we analyze quantum quenches where meson conservation is not a good approximation any longer. This can be achieved, for example, by tuning the post quench transverse field closer to the critical point, thus reducing the fermionic mass and enhancing the Schwinger mechanism. It is worth emphasizing that an efficient representation of $N_{\mathrm{mes}}$ in terms of a MPO is no longer possible as $f_{1}(\ell), f_{2}(\ell)$ show deviations from an exponential decay for values of $h_{\perp}\to1$. The meson number can, nonetheless, be computed by evaluating the terms contained in \eq{eq:Meson-Number-Op final} individually and truncating the sum at large enough $\ell_{\mathrm{max}}$. With this, we indeed observe that the difference between the numerical data and the analytic prediction grows with time as the post quench transverse field is tuned sufficiently close to $1$, in agreement with the observations of Ref. \cite{Lagnese2021}. We, moreover, want to emphasize that results are converged with bond dimension $\chi$, as illustrated in \fig{fig:BreakdownNumberConservation}. Smaller bond dimensions can lead to deviations of the time traces. Ensuring convergence of tensor network results for different choices of $\chi$ is hence crucial to estimate the actual relevance of meson number changing processes.  

We notice the meson number decreases. Hence, rather than the usual Schwinger effect where a large meson decays in two (or more) smaller entities, what dominates the dynamics is the opposite process, namely inelastic scattering of two measons that fuse and become a larger (i.e. more energetic) particle.
\ \\ \ \\
\textbf{Exact diagonalization within the few fermions sector. }
While tensor networks are numerically-exact methods, their applicability is constrained to short times by the entanglement growth, hence they cannot explore the prethermal regime. To overcome this restriction, we neglect the Schwinger mechanism and promote the number of mesons to an exact conservation law, thus projecting the dynamics within a sector with a fixed number of fermions. Furthermore, we wish to focus on the regime of a small transverse field where fermions are well approximated by domain walls.
Hence, we consider the restricted Hilbert space $|j_1,j_2,...j_{2n-1},j_{2n}\rangle=|\uparrow_1...\uparrow_{j_1-1}\downarrow_{j_1}....\rangle$ generated by all the states with $n$ mesons, with ordered coordinates $j_{i+1}-j_i>1$ and having values on the interval $[1,L]$. While the full Hilbert space in the spin basis grows as $2^L$, the restricted Hilbert space grows polynomially $\simeq \frac{1}{(2n)!}L^{2n}$ and much larger system sizes can be reached. This allows us to approach the regime where mesons are well separated, i.e., where our thermodynamic assumptions are valid. The same regime is naturally  obtained after a quantum quench. By further taking into account translational invariance, the exponent of the polynomial growth in $L$ can be lowered by one unit, allowing us to simulate the dynamics of three mesons on $L=100$ for very long times and eventually observing prethermalization (see Fig. \ref{fig_ED}).
Further technical details on this method and benchmarks are discussed in supplementary information \cite{Note1}.

\ \\
\textbf{Semiclassical simulations.} 
For large scale simulations in the semiclassical regime, we relied on a Truncated Wigner Approximation \cite{Polkovnikov2010} which consists of the following steps (see also Refs. \cite{Vovrosh2021a,Scopa2021} for similar approximations)
\begin{enumerate}
    \item Approximate the true Hamiltonian with the projected dynamics within the subspace with a fixed number of fermions (and its multiparticle generalization). This assumption is reliable as long as the Schwinger effect can be neglected.
    \item Approximate the quantum evolution with a classical one:
    \begin{enumerate}
        \item Replace quantum expectation values with proper averages over classical ensembles of particles.
        \item Replace the quantum evolution with properly chosen classical equations of motion, derived from the semiclassical Hamiltonian \eqref{eq_cl_ham} (and its multiparticle generalization).
        \item Classical configurations are sampled from the classical statistical ensemble, then deterministically evolved with the equations of motion. The expectation values of observables is recovered by averaging over the initial conditions.
    \end{enumerate}

\end{enumerate}
The Truncated Wigner Approximation is expected to work in the semiclassical regime, i.e., in the case of weak confinement; see below. However, as it is well known in the literature, one should be aware that this is an uncontrolled approximation in the sense that quantum corrections cannot be easily included in the approximation in a systematic way.
    
To further quantify the method and keep the notation under control, we now focus on the dynamics in the two-particle sector: the generalization to the multiparticle case can be directly obtained.
Let us assume in full generality that the initial state is described by a density matrix $\hat{\rho}_{2\text{pt}}$, we focus on the matrix elements in a coordinate representation for the position of the two fermions $|j_1,j_2\rangle$.
The Wigner distribution $W$ is defined through a partial Fourier transform of the matrix elements in the coordinate basis

\begin{multline}
\label{S_1wig}
\langle x_1+y_1/2,x_2+y_2/2|\hat{\rho}_{\text{2pt}}| x_1-y_1/2,x_2-y_2/2\rangle=\\
\int \dd k_1\dd k_2\,  W(x_1,k_1,x_2,k_2) e^{i y_1 k_1+i y_2 k_2}.
\end{multline}
Above, one should impose integer values of the coordinates, but this restriction will not be important since classical physics emerges in the regime where the matrix elements are smooth functions of the coordinates, hence the discreteness of the lattice becomes irrelevant.
We now move on to consider the dynamics by computing the Heisenberg equation of motion $i\partial_t \hat{\rho}_{\text{2pt}}=[\hat{H}_{\text{2pt}},\hat{\rho}_{\text{2pt}}]$, where $\hat{H}_{\text{2pt}}$ is the quantum Hamiltonian projected in the two-fermions sector, namely the quantized version of Eq. \eqref{eq_cl_ham}. When expressing the Heisenberg equation of motion in terms of the Wigner distribution, one obtains after some straightforward calculations (we omit the $W-$arguments for the sake of notation)

\begin{multline}
\label{S_eqwig}
\partial_t W+[v(k_1)\partial_{x_1}+v(k_2)\partial_{x_2}]W\\
-V'(x_1-x_2)\left(\partial_{k_1}-\partial_{k_2}\right)W\simeq 0
\end{multline}
where $v(k)=\partial_k\epsilon(k)$ and $V'(x)=\partial_x V(x)$ with $V(x)=2h_{\parallel}\bar{\sigma}|x|$.
The above equation is nothing else than the classical Liouville equation for the phase space distribution $W(x_1,k_1,x_2,k_2)$ evolving with the classical Hamiltonian $\mathcal{H}=\epsilon(k_1)+\epsilon(k_2)+2\hpa\bar{\sigma}|x_1-x_2|$.
In the derivation, one assumes that both the matrix element and the potential $V(x)$ are sufficiently smooth in the coordinates. Contributions neglected in the above equation are further orders in the derivative expansion. While $V(x)$ is not strictly speaking smooth, in the limit of weak longitudinal field the cusp in $V(x)$ gives negligible contributions. Notice that, if $h_{\parallel}$ is weak, smooth Wigner distributions will remain smooth during the evolution, ensuring the consistency of the approximation.

We finally turn to the problem of determining the initial Wigner distribution resulting from the quench protocol. 
To this end, we can resort to the quasiparticle picture of quantum quenches in the Ising chain \cite{Calabrese2007}, where the initial state is regarded as an incoherent gas of pairs of particles with opposite momentum $(k,-k)$, the probabilty distribution $n(k)$ of the pair can be computed from the exact solution of the quench in the transverse field \cite{Calabrese2011} (see Ref. \cite{Scopa2021} and supplementary information \cite{Note1} for details and corrections due to the finite longitudinal field).
The distribution $n(k)$ fixes the probability distribution of a single pair of fermions: since pairs are independently created in a homogeneous fashion, we impose that pairs are distributed according to a Poisson distribution.

\ \\{\large\textbf{Data and code availability}} \\Data analysis and simulation codes are available on Zenodo upon reasonable request~\footnote{All data and simulation codes are available upon reasonable request at \url{10.5281/zenodo.7034368}}.


\ \\{\large\textbf{Acknowledgements}}\\
We thank S. Scopa and P. Calabrese for collaboration on closely related topics and A. Lerose for useful discussions. We acknowledge support from the Deutsche Forschungsgemeinschaft (DFG, German Research Foundation) under Germany’s Excellence Strategy--EXC--2111--390814868, TRR80 and DFG grants No. KN1254/1-2 and No. KN1254/2-1, the European Research Council (ERC) under the European Union’s Horizon 2020 research and innovation programme (grant agreement No. 851161), as well as the Munich Quantum Valley, which is supported by the Bavarian state government with funds from the Hightech Agenda Bayern Plus. \\

\ \\{\large\textbf{Author Contributions}}\\
A.B. and S.B. carried out the numerical simulations contained in this work. M.K. supervised the work. All authors contributed critically to the writing of the manuscript and the interpretation of numerical and analytical results.

\ \\{\large\textbf{Competing Interests}}\\
The authors declare no competing interests.

\onecolumngrid
\newpage

\begin{center}
{\large Supplementary Information \\ 
\titleinfo
}\\
Stefan Birnkammer, Alvise Bastianello, Michael Knap
\end{center}

\tableofcontents

\bigskip 
\bigskip


\section{The weakly confined transverse Ising chain}
\label{SM_sec_conf_is}

For the sake of completeness, in this section we summarize the basics of the Ising spin chain in a weak longitudinal field, discussing the dynamics projected in the subspace with a fixed number of fermions and the quench dynamics in the protocol of interest.
\bigskip

\emph{The Ising chain in pure transverse field ---} As a starting point, we consider first the Ising chain in a pure transverse field ($\hpa=0$ in Eq.~(1)). This is a well-known exactly solvable model equivalent to free fermions, see e.g. Ref. \cite{Calabrese2011} for a extensive discussion. Spinless fermions $\{\hat{c}_i,\hat{c}^\dagger_j\}=\delta_{i,j}$ are defined through a Jordan-Wigner transformation
\be
\frac{\hat\sigma_j^x+i\hat\sigma_j^y}{2}=\exp\big(i\pi \sum_{i<j} \hat{c}^\dagger_{i}\hat{c}_{i}\big)\hat{c}^\dagger_j\,. 
\ee
The Ising Hamiltonian commutes with the parity operator $\hat{P}=\prod_j \hat\sigma_j^x$, thus the Hilbert space splits into two disconnected parts of different parity $\hat{P}=\pm 1$: the spinless fermions obey different boundary conditions depending on the parity sector, but this subtlety can be ignored in the infinite volume limit.
The transverse Ising Hamiltonian is then diagonalized in the Fourier basis after a Bogoliubov rotation
\be\label{S_eq_bog_rot}
\begin{pmatrix}\hat{c}_j \\[4pt] \hat{c}^\dagger_j \end{pmatrix}=\int \frac{\dd k}{\sqrt{2\pi}} e^{i k j}\begin{pmatrix} \hat\alpha(k) \\[4pt] \hat\alpha^\dagger(-k)\end{pmatrix}=\int \frac{\dd k}{\sqrt{2\pi}} e^{i k j}U_{\theta_k}\begin{pmatrix} \hat\gamma(k) \\[4pt] \hat\gamma^\dagger(-k)\end{pmatrix}\, , \hspace{2pc} U_{\theta_k}=\begin{pmatrix} \cos\theta_k && i\sin\theta_k \\ i\sin\theta_k && \cos\theta_k\end{pmatrix}\,,
\ee
where the modes $\hat{\gamma}(k)$ obey the canonical anticommutation rules $\{\hat{\gamma}(k),\hat{\gamma}^\dagger(q)\}=\delta(k-q)$. The angle $\theta_k$ parametrizes the Bogoliubov rotation and the choice
\be\label{Seq_Bog_angle}
\theta_k=-\frac{1}{2i}\log\left(\frac{h_\perp-e^{i k}}{(\cos k-h_\perp)^2+\sin^2k}\right)\, 
\ee
diagonalizes the Hamiltonian $\hat{H}=\int \dd k\, \epsilon(k) \hat{\gamma}^\dagger(k)\hat{\gamma}(k) +\text{const.}$ with $\epsilon(k)=2\sqrt{(\cos k-\hpe)^2+\sin^2k}$. In the fermion language, the Hilbert space can be described as a Fock space built by acting with the creation operators on the vacuum $|0\rangle$ (defined as $\hat{\gamma}(k)|0\rangle=0$), which is also identified with the ground state of the chain.
Homogeneous quantum quenches in the transverse Ising has been first addressed in Ref. \cite{Calabrese2011}: in this framework, one initializes the system in the ground state for a certain transverse field $\tilde{h}_{\perp}$ and brings the system out of equilibrium by changing $\tilde{h}_{\perp}\to \hpe$. By using that the initial state is annihilated by the modes that diagonalize the prequench Hamiltonian and that the pre- and post-quench modes are connected through a Bogoliubov rotation, one can write the initial state in a simple squeezed form
\be\label{Seq_initial_state}
|\psi\rangle\propto \exp\left[\int_0^\pi \dd k\, K(k) \hat{\gamma}^\dagger(k)\hat{\gamma}^\dagger(-k)\right]|0\rangle\, ,
\ee
with the wavefunction $K(k)$ determined by the difference of the pre and post quench Bogoliubov angles $K(k)=-i \tan(\theta_k^\text{post}-\theta_k^\text{pre})$. By similar means, in Section \ref{S_sec_stagg} we will determine the initial state obtained after a staggered pulse in the transverse magnetization.

\bigskip

\emph{Effects of weak confinement: the projected dynamics ---} We now consider the activation of a non-trivial longitudinal field $\hpa$. Following Ref. \cite{Rutkevich2008}, one expresses the full Hamiltonian in the basis of the modes of the transverse part. Hence, let $|\{k_i\}_{i=1}^N\rangle$ be a multi-fermionic state: apart from a non-essential constant offset, the Hamiltonian reads
\be\label{S_eq_fermionH}
\hat{H}|\{k_i\}_{i=1}^N\rangle= \left(\sum_{i=1}^N \epsilon(k_i)\right)|\{k_i\}_{i=1}^N\rangle+\hpa\sum_{M=1}^\infty \frac{1}{M!}\int_{-\pi}^\pi \frac{\dd^M q}{(2\pi)^M}\, 2\pi\delta\left(\sum_{j=1}^M q_j-\sum_{i=1}^N k_i\right) |\{q_j\}_{j=1}^M\rangle \langle\{q_j\}_{j=1}^M|\hat\sigma^z_0|\{k_i\}_{i=1}^N\rangle\, .
\ee
The matrix elements of the longitudinal magnetization $ \langle\{q_j\}_{j=1}^M|\hat\sigma^z_0|\{k_i\}_{i=1}^N\rangle$ ---also called form factors--- are known and, by means of a repeated use of the Wick theorem, are entirely encoded in the two-fermion form factors $\langle k_1 k_2|\hat\sigma^z_0|0\rangle$, $\langle 0|\hat\sigma^z_0|k_1, k_2\rangle$ and $\langle k_1|\sigma_0^z|k_2\rangle$, see Ref. \cite{Rutkevich2008} for further details.
The Hamiltonian in \eq{S_eq_fermionH} is valid for arbitrary values of the longitudinal field $\hpa$. However, in the limit of weak confinement $|\hpa|\ll \min_k \epsilon(k)$, important simplifications can be invoked.
As we mentioned in the main text, the effective dynamics conserves the number of fermions on very large time scales, hence we can project the dynamics of \eq{S_eq_fermionH} in the number-conserving space.
Secondly, the interaction in the number-conserving sector splits into a long-range linear potential (inducing the confinement) plus short range corrections \cite{Rutkevich2008} (see also Ref. \cite{Scopa2021}). Short range corrections are negligible when compared with the linear term (and are furthermore linearly suppressed for $\hpe$ small). With these considerations, in the two particle sector one reaches the simple Hamiltonian
\be\label{eq_2pt_H}
\hat{H}_\text{2pt}= \int \dd k_1\dd k_2 [\epsilon(k_1)+\epsilon(k_2)] |k_1,k_2\rangle \langle k_1,k_2|+\sum_{j_1,j_2}  2\hpa\bar{\sigma} |j_1-j_2| |j_1,j_2\rangle\langle j_1 j_2|\, ,
\ee
where the potential term is more conveniently expressed in the real basis $|j_1,j_2\rangle\equiv\int \frac{\dd k_1\dd k_2}{2\pi} e^{i k j_1+ik_2 j_2}|k_1,k_2\rangle$. The expression can be directly generalized beyond the two fermions sector.
We notice that for small transverse fields, the kinetic energy reads $\epsilon(k)\simeq 2-2\hpe\cos(k)+...$ and the kinetic term becomes a simple nearest-neighbor hopping.
For the sake of simplicity, in Fig.~2 of the main text we consider this limit, where in addition fermions can be identified with sharp domain walls in the Ising chain. The more general case of finite $\hpe$ can be also addressed in a similar way.
When considering the quench protocol of interest, the state of \eq{Seq_initial_state} can be used as the initial condition of the dynamics: this proposal has been originally made in Ref. \cite{Kormos2017}. However, while at late time the Schwinger mechanism is extremely suppressed, it has been recently understood that the activation of a longitudinal field non-trivially affects the number of fermions \cite{Scopa2021}. This additional production of excitations ceases after a short time transient and effectively renormalizes the wavefunction
\be
K(k)\to \mathcal{K}(k)=-i \tan(\theta_k^\text{post}-\theta_k^\text{pre})-i \hpa \bar{\sigma} v(k)/\epsilon^2(k)\, ,
\ee
with $v(k)=\partial_k \epsilon(k)$ being the group velocity. While this renormalization can be important to check meson conservation for relatively strong longitudinal field (eg Figs.~1(c)), its effect is negligible for the parameter choice of Figs.~3 and~4 and it is thus neglected.

\bigskip
\emph{Quantization of the mesonic states ---} 
Mesons have internal degrees of freedom associated with the different energy levels of the two-particle problem described by $\hat{H}_{2\text{pt}}$ \eq{eq_2pt_H}. In the limit of weak transverse field  $\epsilon(k)\simeq 2-2\hpe\cos(k)+\mathcal{O}(\hpe^2)$ the eigenfunctions and energies can be exactly computed by means of Bessel functions \cite{Rutkevich2008}. For finite transverse field, a simple analytical solution is not available, but $\hat{H}_{2\text{pt}}$ can be easily numerically diagonalized (see eg. Ref. \cite{Scopa2021}).
Nonetheless, the limit where the longitudinal field is much weaker than the transverse one $\hpa\ll \hpe$ is amenable of semiclassical methods. Already in the seminal paper \cite{Kormos2017}, the semiclassical quantization of meson energies has been observed to be in very good agreement with numerical data, even far from the extreme limit $\hpa\ll \hpe$. Therefore, further motivated by our goal to describe the prethermal phase observed in the classical regime, we briefly review the semiclassical quantization of mesonic masses \cite{Rutkevich2008}: the pure classical limit is then recovered in the limit of vanishing longitudinal field, where the quantized energies merge in a continuum.

In order to find the momentum-dependent energy levels $\{\mathcal{E}(J,k)\}_{J}$, let us address the classical two-fermion problem associated with \eq{eq_2pt_H}. It is convenient to consider the center-of-mass and the relative coordinates, $(X, k)$ and $(x, q)$ respectively.
For this choice of coordinates the $2-$particle classical Hamiltonian takes the simple form
\be
\label{eq:H_rel}
        H_{2\text{pt}}(k, q, X, x) = \epsilon( k/2+q) + \epsilon(k/2-q) + \chi  \abs{x} \equiv \omega(q, k) + \chi \abs{x}
\ee
where we introduced the notation $\chi=2\hpa\bar{\sigma}$ and $\omega(k, q)=\epsilon(k/2+q) + \epsilon(k/2-q)$. The total momentum $k$ is conserved.

Generally $q(t)$ and $x(t)$ describe an oscillating motion associated to the breathing of the fermion bound state. The latter is formally captured by 
\begin{align}
\label{eq:Classical_q_x}
    &q(t)= \begin{cases} \begin{aligned}&
    q_{a}(E, k) - \chi t &\quad t \in [0, t_{1}]\\
    &q_{b}(E, k) + \chi (t - t_{1}) &\quad t \in [t_{1}, 2t_{1}]
    \end{aligned} \end{cases} \qquad &x(t)=\begin{cases}\begin{aligned}
    &\chi^{-1} \big(E - \omega(k, q(t))\big)  \quad t \in [0, t_{1}]\\
    -&\chi^{-1} \big(E - \omega(k, q(t))\big)   \quad t \in [t_{1}, 2t_{1}]
    \end{aligned} \end{cases}
\end{align}

Here we introduced the points $(q_{a}, q_{b})$ as the turning points of the classical problem, where the total energy of the relative motion is stored in kinetic energy, i.e. the relative distance of the fermions vanishes. These turning points are reached after multiples of the time period $t_{1}=(q_{a} - q_{b})/\chi$. Depending on the choice for $k$ the function $\omega(k,q)$ will either resemble the shape of a single or double-well potential in $q$. This entails two fundamentally different cases for the turning points $q_{a}, q_{b}$. For small values of $k$, $\omega(k,q)$ looks like a single well potential symmetric in $q\to -q$. The turning points will consequentially share this reflection symmetry around $q=0$ and satisfy the constraint $q_{b}=-q_{a}$ independent of the choice of $E$. For larger values of $k$ we, however, find a different behavior. In this case $\omega(k,q)$ can be interpreted as a double-well potential and the motion can be stuck within a single-well for sufficiently small energies $E<\omega(k, q=0)$. In consequence, the relation $q_{b}=-q_{a}$ is no longer valid and the motion will no longer be symmetric under the reflection $q\to -q$.\\ 
In order to determine the mesonic energy bands $\mathcal{E}(J,k)$ we apply a semiclassical Bohr-Sommerfeld quantization to the set of conjugate variables $(q(t), x(t))$ given by
\be
\label{eq:Bohr-Sommerfeld-Quantization}
  J=  \oint \meas{x}\;q(x) = 2\pi \big(n - \frac{1}{2}\big) \qquad \text{with} \quad n \in \mathbb{N}.
\ee
Using the functional form of $q(t)$ and $x(t)$ of \eqw{eq:Classical_q_x} we thus find
\be
\label{eq:QuantizationDisperionsBands}
 J=2\chi^{-1}   \mathcal{E}(J,k) (q_{b}-q_{a}) - 2\chi^{-1}\int_{q_{a}}^{q_{b}} \meas{q}\;\omega(k, q)   
\ee
However, due to the Pauli exclusion principle $n$ is forced to be even when the two fermions can come in contact.
We thus find two different conditions referring to the cases of a symmetric motion ($n$ even integer) and a motion stuck within a single well of the double-well potential ($n$ integer).
For a more detailed derivation of this quantization procedure we refer to Ref.~\cite{Rutkevich2008}. Solving \eq{eq:QuantizationDisperionsBands} numerically equips us with all information required to describe the thermodynamics of the prethermal state, as we now discuss.

\section{Thermodynamics for mesonic systems}
\label{sec_thermodynamics}

In the extremely dilute regime, the typical size of a meson is negligible with respect to their relative distance and their thermodynamics can be approximated as if they were point-like particles. However, mesons are extended objects and their size gives non-neglible contributions moving aside from the extreme dilute scenario.
Therefore, we now aim to a better treatment where mesons are considered as extended objects with a fixed length $\ell(J,k)$ which in first approximation can be taken as the average magnetization. Despite the fact this is a rather crude approximation (for example, the size of the meson oscillates in time), it nicely captures features beyond the extreme dilute scenario. 
Hence, we now discuss the thermodynamics of a gas of hard-rods, where the length of the meson depends on its momentum $k$ and internal energy level $J$. The momentum $k$ is quantized in units of $2\pi/L$, but we are eventually interested in the thermodynamic limit and replace summations with integrals. Even though we wish to address directly the semiclassical limit, considering the correct momentum quantization is necessary to obtain the correct phase-space normalization of the thermal curves.
In the limit of a weakly interacting many-meson system all thermodynamic information is contained in the grand-canonical partition function

\be
\label{eq:PartitionFunction}
\mathcal{Z} = \sum_{\{\rho_J(k)\}_{J,k}}e^{\mathcal{S}}e^{-L\beta \sum_{J,k} \rho_J(k)(\mathcal{E}(J,k) -\mu)}\,,
\ee
where $\rho_J(k)$ is the density of mesons with quantum numbers $(J,k)$ and energies $\mathcal{E}(J,k)$. In this notation, $J$ is the classical action variable that is discretized in units of $2\pi$ in the quantum case, due to the Bohr-Sommerfeld quantization condition.
The summation over the densities should be interpreted in a path integral sense, but it is useful to think about it as a discrete object first and take the limit at the end.
The system  size is $L$.
We also introduce the number of mesons with quantum number $(J,k)$ as $N(J,k)=L \dd k \rho_J(k)$, with $\dd k$ the size of a small momentum cell.

The chemical potential $\mu$ ensures the meson number conservation present in the prethermal regime.
The summation included in \eq{eq:PartitionFunction} takes into account all possible ways to distribute a given number of mesons in the system.
While the summation over $\{\rho_J(k)\}_{J,k}$ spans the possible populations for each quantum number, the entropic factor $e^{\mathcal{S}}$ counts the possible spatial arrangements of the mesons for a given population distribution.
The entropic term is sensitive to the length of the mesons: as anticipated, let $\ell(J,k)$ be the effective length of a meson, which we approximate as a constant. To simplify notation, we group the two quantum numbers in a single one $(J,k)\to \eta$, hence  $\ell(J,k)\to \ell_\eta$.
This identification takes our system of mesons to be a system of hard-rods of different lengths $\{\ell_\eta\}_{\eta}$ with $N_\eta$ particles for each species.
For the sake of simplicity, we now choose to neglect the discreteness of the lattice and treat the position of a meson as a continuous variable. This approximation is valid either in the small density limit where finite-volume effects are negligible, or in the limit where mesons are large compared with the unit cell. Therefore, we expect corrections can be important only in the extreme limit of very dense and tight mesons.

To compute the partition function, we first introduce an ordering in the particle species: mesons with quantum number $\eta$ will be contained in an interval of length $L_\eta$, where $L=\sum_\eta L_\eta$.
The ordered partition function due to the spatial degrees of freedom is readily computed as (we assume $\eta$ runs from $1$ to $m$)
\begin{align}
\label{eq:PartitionFunction-TonksN-ordered}
    \mathcal{Z}_{m}^{(ord)} &= \int_{0}^{L}\meas{L_{1}}\meas{L_{2}}\dots\meas{L_{m}}\;\delta\Big(L-\sum_{i}L_{i}\Big) \Prod_{r=1}^{m} \mathcal{Z}_{1}(L_{r},N_{r})
    = \int_{L_{1}^{(min)}}^{L_{1}^{(max)}}\meas{L_{1}}\int_{L_{2}^{(min)}}^{L_{2}^{(max)}}\meas{L_{2}}\dots \int_{L_{m}^{(min)}}^{L_{m}^{(max)}}\;  \Prod_{r=1}^{m} \mathcal{Z}_{1}(L_{r},N_{r}) \nonumber\\
    &= \frac{1}{(N_{1}+N_{2}+\dots+N_{m})!} \Prod_{r=1}^{m} (L - N_{1}\ell_{1} - N_{2}\ell_{2} - \dots - N_{m}\ell_{m})^{N_{r}}.
\end{align}
Here we introduced the minimal and maximal length of subsystem $r$ as $L_{r}^{(min)}=N_{r}\ell_{r}$ and $L_{r}^{(max)}= L - \sum_{n<r}L_{n} - \sum_{n>r} N_{n}\ell_{n}$, respectively, to make the constraint of the $\delta$-distribution explicit. The first, hereby, just takes into account that the subsystem $r$ containing $N_{r}$ particles of length $\ell_{r}$ can not be compressed further than $L_{r}^{(min)}$. The upper bound results from the opposite situation. Having already fixed the subsystem sizes $L_{1},...,L_{r-1}$ we can compress the remaining subsystems to their minimal configuration and find $L_{r}^{(max)}$. For the last equality of \eq{eq:PartitionFunction-TonksN-ordered} we used the well-known result for the partition function of a simple gas of a single species of hard-rods $\mathcal{Z}_{1}(L,N)=[(L - N\ell)]^{N}/N!$. 
Since the different particles are grouped into subsystems the partition function of \eq{eq:PartitionFunction-TonksN-ordered} does not take into account reordering of the individual meson species. This can, nevertheless, be accounted for by a combinatorial factor 
\begin{figure}[t!]
    \centering
    \includegraphics[width=\textwidth]{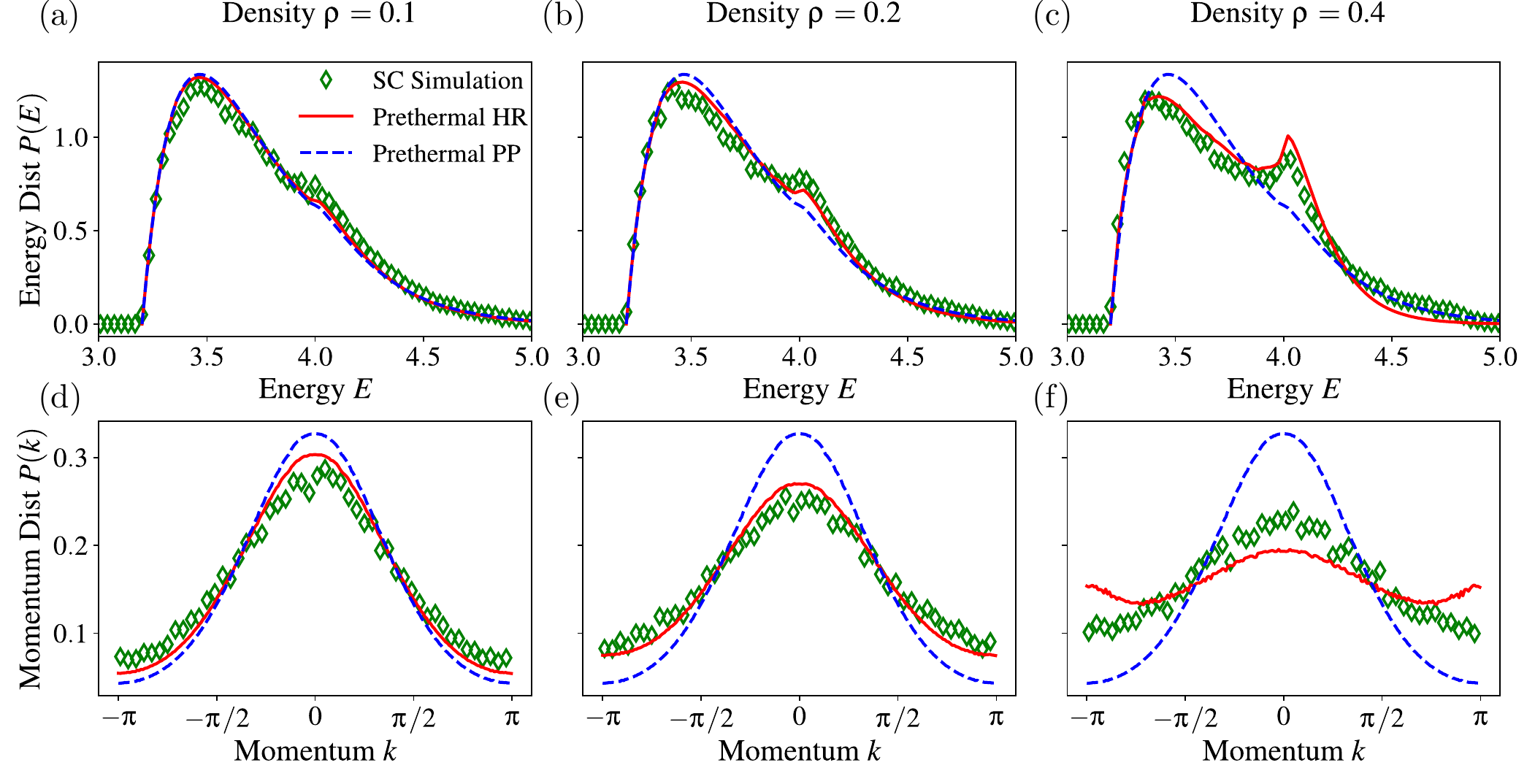}
    \caption{\textbf{Effect of finite-volume corrections.} We benchmark our predictions from thermodynamic analysis for a system of point particles (PP) as well as for an ensemble of hard-rods (HR) against the results of semiclassical simulations. (a)-(c) We display the energy distribution $P(E)$ of the prethermal steady state for different densities of mesons $\rho$ in a system characterized by a longitudinal and transverse field $(\hpa, \hpe)=(0.015, 0.2)$. The initial energy per meson is thereby on average given by $E=3.73737$. We find that the finite-volume correction included in the HR treatment (red solid curves) improves the thermodynamic description of the semiclassical data points (green markers) compared to the PP description (blue dashed curves). (d)-(f) The prethermal state of hard rods captures the momentum distribution $P(k)$ especially well for densities $\rho\lesssim0.2$. For very high densities $\rho\approx0.4$ deviations become apparent.}
    \label{fig:Fig7-ZeroFiniteVolComparison}
\end{figure}
\begin{align}
\label{eq:PartitionFunction-TonksN}
    \mathcal{Z}_{m} &= \binom{N_{1} + N_{2} + \dots + N_{m}}{N_{m}} \binom{N_{1} + N_{2} + \dots + N_{m-1}}{N_{m-1}}\dots \binom{N_{1} + N_{2}}{N_{2}}\;\mathcal{Z}_{m}^{(ord)} \nonumber \\
    &= \Prod_{r=1}^{m}\Bigg(\frac{1}{ N_{r}!}\Bigg) \; (L - N_{1}\ell_{1} - N_{2}\ell_{2} - \dots - N_{m}\ell_{m})^{N_{1} + N_{2} + \dots + N_{m}}. 
\end{align}

When computing \eq{eq:PartitionFunction-TonksN-ordered}, we did not take into account the momentum quantization which allows for a reshuffle of the particles within the single momentum cell of width $\dd k$. Due to quantization, such a cell has $\frac{L}{2\pi}\dd k$ available quantum numbers. We now consider the problem of arranging $N_r$ indistinguishable particles on $\frac{L}{2\pi}\dd k$ sites. We neglect the case of double occupancies, which is equivalent to consider a classical statistics (in contrast with Fermi or Bose statistics). In this case, the number of possible arrangements is $\frac{1}{N_r!}\left(\frac{L\dd k}{2\pi}\right)^{N_r}=\frac{L^{N_r}}{N_r!}\left(\frac{\dd k}{2\pi}\right)^{N_r}$. Now, we argue that the prefactor $L^{N_r}$ has already been taken into account in \eq{eq:PartitionFunction-TonksN-ordered}, while the factor $1/N_r!$ has already been accounted when considering \eq{eq:PartitionFunction-TonksN}. Hence, one is left with the additional phase-space contribution $\left(\frac{\dd k}{2\pi}\right)^{N_r}$.
We can now finally identify the entropic term as
\be
\mathcal{S}=\log\left(\mathcal{Z}_m\prod_r \left(\frac{\dd k}{2\pi}\right)^{N_r}\right)
\ee
In the hypothesis that $N_r$ is large and using a Stirling approximation for the factorial, one finds
\be
    \mathcal{S}/L \approx \Sum_{r=1}^{m} \dd k\rho_{r}\Big[ \ln\Big(\frac{1 - \dd k\rho_{1}\ell_{1} - \dd k\rho_{2}\ell_{2} -\dots - \dd k\rho_{m}\ell_{m}}{2\pi\rho_{r}}\Big) + 1\Big].
\ee
By the extensivity of the entropic term, in the thermodynamic limit $L\to\infty$ the partition function localizes to the saddle point of the free energy $\mathcal{F}=-\beta^{-1} \log\mathcal{Z}$
\be
-L^{-1}\beta\mathcal{F}=\sum_J\int \dd k\rho_{J}(k)\Big[ \ln\left(1 - \sum_{J'}\int\dd q \ell(J',q)\rho_{J'}(q)\right)-\ln(2\pi\rho_{J}(k)) + 1-\beta(\mathcal{E}(J,k)-\mu)\Big]
\ee
Finally, imposing $\delta \mathcal{F}/\delta \rho_J(k)=0$ we find
\be\label{eq_S_rhonk}
\rho_J(k)=\frac{1-\rho M}{2\pi}\exp\left[-\beta(\mathcal{E}(J,k)-\mu)-\frac{\rho \ell(J,k)}{1-\rho M}\right]
\ee
Where the meson coverage $\rho M$, the total density $\rho$ and the mean energy $E$ are
\begin{align}
    \label{eq:Density-Constraint}
    &\text{(i)} & \rho &=\sum_J \int \dd k\, \rho_J(k)  &\\
    \label{eq:Energy-Constraint}
    &\text{(ii)} & E  &=\sum_J \int \dd k\, \mathcal{E}(J,k)\rho_J(k) &\\
    \label{eq:Selfconsistency-Constraint}
    &\text{(iii)} & \rho M &=\sum_J \int \dd k\, \ell(J.k) \rho_J(k).  &
\end{align}
The meson density and coverage must be computed self consistently with \eq{eq_S_rhonk}.
On prethermal states, thus enforcing the meson density conservation, the chemical potential $\mu$ and inverse temperature $\beta$ must be chosen to match the initial density $\rho$ and energy $E$. On the other hand, thermal ensembles fix only the energy $E$ and $\mu=0$.
It is worth to consider the momentum-energy probability distribution of a meson $P(E,k)$ in the classical case. We start considering $P(E,k)$ in the quantum regime writing it as
\be
P(E,k)=\sum_J \rho_J(k)\delta(E-\mathcal{E}(J,k))\, ,
\ee
with $\delta$ a Dirac delta. In the semiclassical regime, we can replace the summation over $J$ with an integral $\sum_J \to \int \frac{\dd J}{2\pi}$, where the factor $2\pi$ comes from the Born-Sommerfeld quantization condition $J=2\pi(n-1/2)$
\be
P(E,k)=\frac{1-\rho M}{(2\pi)^2}e^{-\beta (E-\mu)}\int \dd J
\delta(E-\mathcal{E}(J,k)) e^{-\frac{\rho \ell(J,k)}{1-\rho M}}\, .
\ee
Notice that in the small density limit $\rho\to 0$, the coverage $\rho M$ vanishes as well and one recovers the usual thermal distribution of non-interacting particles.

Finally, we wish to explicitly discuss how $\ell(J, k)$ is computed: as we already mention, we estimate the length of the hard rod approximation of the meson with its average length. In the classical limit, this amounts to computing the time average of the distance between the two fermions within one period of the oscillation. In the quantum regime, instead, $\ell(J,k)$ is computed by considering the quantum expectation value of the relative distance $|x|$ (see \eq{eq:H_rel}) on the energy eigenstate. Of course, the two definitions coincide in the semiclassical limit.

To further emphasize the importance of the finite volume corrections, we refer to \fig{fig:Fig7-ZeroFiniteVolComparison} providing a benchmark of our thermodynamic predictions against semiclassical simulations. The semiclassical data for the energy distribution $P(E)$ of the prethermal steady state shown in \figcs{fig:Fig7-ZeroFiniteVolComparison}{a}{c} reveals that systems of higher meson densities $\rho$ in fact exhibit increased meson occupancy towards the edges of the Brillouin zone ($E\approx4.0$), where mesons show their smallest average lengths. While the conventional thermodynamic ansatz for point particles fails in reproducing this feature of the steady state, including finite volume corrections indeed allows us to capture this effect. This is further supported by results for the momentum distributions $P(k)$ of the same systems, as illustrated in \figcs{fig:Fig7-ZeroFiniteVolComparison}{d}{f}.

\section{Initializing moving mesons by modulated pulses of the transverse field}
\label{S_sec_stagg}

The prethermalization time scale can be strongly reduced when mesons are initialized with a finite velocity, as we demonstrated in Fig.~2. This scenario can be achieved through a different state preparation, by replacing the homogeneous quench in the transverse field with a modulated pulse on a period of $n$ sites \cite{Bastianello2018}, i.e., $\hat{H}_\text{Pulse}=-\delta(t) \sum_j \rh_j \sigma^x_j$, where $\rh_{j+n}=\rh_j$.
While fermions are still excited in pairs $(k,k')$, the periodic modulation breaks translation invariance thus giving a non-trivial momentum to the meson $k+k'=2\pi j/n$, with $j$ an integer.
While this strategy allows us to maintain the dilute meson approximation, it creates moving mesons right from the beginning, thus promoting scrambling.
In particular, let us assume the state is initially prepared in the ground state $|0\rangle$ of the transverse-field Ising model $\hat{H}_\text{tr}=-\sum_{j}\sigma_{j+1}^z\sigma_j^z+\hpe \sigma_j^x$, then we consider a pulse Hamiltonian in the form $\hat{H}_\text{Pulse}=-\delta(t) \sum_j \rh_j \sigma^x_j$, with $\rh_j$ being a periodic modulation with period $n$, i.e.  $\rh_j=\rh_{j+n}$.
After the pulse application, the state evolved into $|\psi\rangle\equiv e^{i\sum_j \rh_j \sigma_j^x}|0\rangle$ which we now characterize. 
After the pulse, the longitudnal field is activated and $|\psi\rangle$ is used as the initial state for the confining dynamics: we assume the regime of weak confinement, hence we neglect meson production caused by the activation of the longitudinal field $\hpa$.
As a first step, we express the pulse in the fermionic basis and eventually in the modes of the transverse Ising Hamiltonian, by \eq{S_eq_bog_rot}. For the sake of a more compact notation, we define the non-rotated fermionic modes in the Fourier basis as $\hat{\alpha}(k)=\cos\theta_k\hat{\gamma}(k)+i\sin\theta_k \hat{\gamma}^{\dagger}(-k)$, where $\{\hat{\alpha}(k),\hat{\alpha}^\dagger(q)\}=\delta(k-q)$. We write the pulse as
\begin{multline}\label{eq_S8}
-\sum_j \rh_j\sigma_j^x=\sum_j 2\rh_j c^\dagger_{j}c_j+\text{const.}=
\int_{-\pi}^{\pi} \frac{\dd k\dd q}{2\pi} \left[\sum_j 2\rh_j e^{ij(k-q)}\right]\hat{\alpha}^\dagger(q)\hat{\alpha}(k)+\text{const.}=\\
\int_{-\pi}^\pi \dd k\dd q\, 2\tilde{\rh}(k-q)\delta(e^{in (k-q)}-1)\hat{\alpha}^\dagger(q)\hat{\alpha}(k)+\text{const.}=\sum_{j,j'}\int_{-\pi/n}^{\pi/n} \dd k\, \frac{2}{n}\tilde{\rh}(2\pi (j-j'))\hat{\alpha}^\dagger(k+2\pi j'/n)\hat{\alpha}(k+2\pi j/n)+\text{const.}\, .
\end{multline}
Above, we then use the periodicity of $\rh_j$ to extract a delta function in the momentum space and defined $\tilde{\rh}(k)\equiv \sum_{j=0}^{n-1}e^{ik j}\rh_j$.
We now use this expression to show that
\be\label{eq_S9}
|\psi\rangle=e^{i\sum_j \rh_j\sigma_j^x}|0\rangle=\mathcal{N} \exp\left[\frac{1}{2}\sum_{jj'}\int_{-\pi/n}^{\pi/n}\dd k\, \mathcal{M}_{j,j'}(k) \hat{\gamma}^\dagger(k+j2\pi/n)\hat{\gamma}^\dagger(-k-j'2\pi/n)\right]|0\rangle\, .
\ee
Above, $\mathcal{M}_{j,j'}(k)$ is a $k-$dependent $n\times n$ complex matrix. Notice that, due to the fermionic anticommutation relations, it holds $\mathcal{M}_{j,j'}(k)=\mathcal{M}_{n-j',n-j}(-k)$.
\eq{eq_S9} generalizes the state of \eq{Seq_initial_state} to the case of fermions pairwise excited, but with pairs possibly having non-zero total momentum.

To show \eq{eq_S9} it is convenient to introduce a fictitious parameter $\lambda$ and considering $|\psi_{\lambda}\rangle=e^{i\lambda\sum_j \rh_j\sigma_j^x}|0\rangle$, by showing infinitesimal $\lambda-$changes move within the parametrization \eqref{eq_S9} with a $\lambda-$ dependent wavefunction $\mathcal{M}_{jj'}^\lambda(k)$ and normalization $\mathcal{N}_\lambda$.
By introducing the $\lambda-$dependence in \eq{eq_S9} and taking the derivative, one obtains
\be
\partial_\lambda |\psi_\lambda\rangle=\left(\frac{\partial_\lambda \mathcal{N}_\lambda}{\mathcal{N}_\lambda}+\frac{1}{2}\sum_{jj'}\int_{-\pi/n}^{\pi/n}\dd k \partial_\lambda\mathcal{M}^{\lambda}_{j,j'}(k) \hat{\gamma}^\dagger(k+j2\pi/n)\hat{\gamma}^\dagger(-k-j'2\pi/n)\right)|\psi_\lambda\rangle\, .
\ee
On the other hand, by the very definition one has $\partial_\lambda |\psi_\lambda\rangle=i\sum_j \rh_j \sigma_j^x|\psi_\lambda\rangle$.
By repeatedly using the fermionic commutation relation on \eq{eq_S9}, the action of $\hat{\alpha}(k+2\pi j/n)$ on $|\psi_\lambda\rangle$ is computed as
\be
\hat{\alpha}(k+2\pi j/n)|\psi_\lambda\rangle=\left[\sum_{s}\cos\theta_{k+2\pi j/n}\mathcal{M}^\lambda_{j,s}(k)\hat{\gamma}^\dagger(-k-s2\pi/n)+i\sin\theta_{k+2\pi j/n}\hat{\gamma}^\dagger(-k-2\pi j/n)\right]|\psi_\lambda\rangle\, .
\ee
Further applying $\hat{\alpha}^\dagger(k+2\pi j'/n)$ on the so-obtained state, one can proceed by the same methods and the action of \eq{eq_S8} on $|\psi_\lambda\rangle$ is easily obtained. After some simple, but tedious calculations it is shown that the gaussian form of \eq{eq_S9} closes the equations if $\mathcal{M}_{j,j'}^\lambda$ satisfies
\begin{multline}\label{eq_S12}
 i\partial_\lambda\mathcal{M}^{\lambda}_{j,j'}(k) =\\
\sum_{s,s'} \frac{2}{n}\tilde{\rh}(2\pi (s-s'))\left[\cos\theta_{k+2\pi s/n}\mathcal{M}^\lambda_{s,j}(k)+i\sin\theta_{k+2\pi s/n}\delta_{j,s}\right]
\left[-i\sin\theta_{k+2\pi s'/n}\mathcal{M}^\lambda_{s',j'}(k)+\cos\theta_{k+2\pi s'/n}\delta_{j',s'}\right]+\\
[k\to -k\,, j\to n-j'\,, j'\to n-j]\, ,
\end{multline}
where the term $[k\to -k\,, j\to n-j'\,, j'\to n-j]$ is obtained from the first through the proper index replacement. Finally, the desired matrix $\mathcal{M}_{j,j'}(k)$ is computed by integrating the above equation up to $\lambda=1$. Albeit a close analytical expression is hard to obtain, the $n\times n$ matrix relation of \eq{eq_S8} can be easily numerically integrated.
Furthermore, in the problem at hand we are mostly interested in the regime of dilute mesons, i.e. when $\rh_j$ is small. \eq{eq_S12} is easily solved at the leading order in $\rh_j$ obtaining

\be
 i\mathcal{M}^{\lambda}_{j,j'}(k) = \frac{2}{n}\tilde{\rh}(2\pi (j-j'))i\sin\theta_{k+2\pi j/n}\cos\theta_{k+2\pi j'/n}+\frac{2}{n}\tilde{\rh}(2\pi (j'-j))i\sin\theta_{-k-2\pi j'/n}\cos\theta_{-k-2\pi j/n}+\mathcal{O}(\rh^2)\, .
\ee

\section{Lattice gauge theories: the example of the $\mathrm{U}(1)$ quantum link model}

As an example how the proposed multistage thermalization mechanism generalizes to lattice gauge theories we want to discuss the physics of a $\mathrm{U}(1)$ quantum link model (QLM)
\begin{equation}
    \label{eq:QLM}
     H_{\mathrm{QLM}}= -\omega \sum_{j=1}^{L-1}(\phi_{j}^{\dagger} S_{j, j+1}^{+}\phi_{j+1}^{\phantom{\dagger}} + \mathrm{h.c.}) + \frac{m}{2} \sum_{j=1}^{L} (-1)^{j} \phi_{j}^{\dagger} \phi_{j}^{\phantom{\dagger}} - 2 h_{\parallel} \sum_{j=1}^{L-1} S_{j, j+1}^{z},
\end{equation}
where staggered Kogut-Susskind fermionic matter~\cite{Kogut1975, Surace2020} is described by the creators $\phi_{j}^{\dagger}$ and annihilators $\phi_{j}^{\phantom{\dagger}}$ and we additionally include gauge degrees of freedom encoded by a local set of spin-$\frac{1}{2}$ operators $\{S_{j, j+1}^{z},S_{j, j+1}^{+},S_{j, j+1}^{-}\}$ acting on the links $(j, j+1)$. The content of fermionic excitations in the model in this staggered convention can be counted using a generalized number operator $n_{j} = \frac{1}{2} [1 - (-1)^{\phi_{j}^{\dagger}\phi_{j}^{\phantom{\dagger}} + j}]$. This relates to the picture that holes located at odd sites can be interpreted as a antiquark ($\bar{q}$), whereas a particle located at a even site resembles a quark ($q$): in the regime where $m$ is infinite, the ground state of the theory is readily identified with the vacuum of the theory $|\text{Vac}\rangle$ (where quarks and antiquark are absent). For large, but finite mass $m$, bare particles must be suitably renormalized, but the picture remains qualitatively the same: for the sake of clarity, we will mainly refer to the large $m$ limit. 

\begin{figure}[t!]
    \centering
    \includegraphics[width=0.95\textwidth]{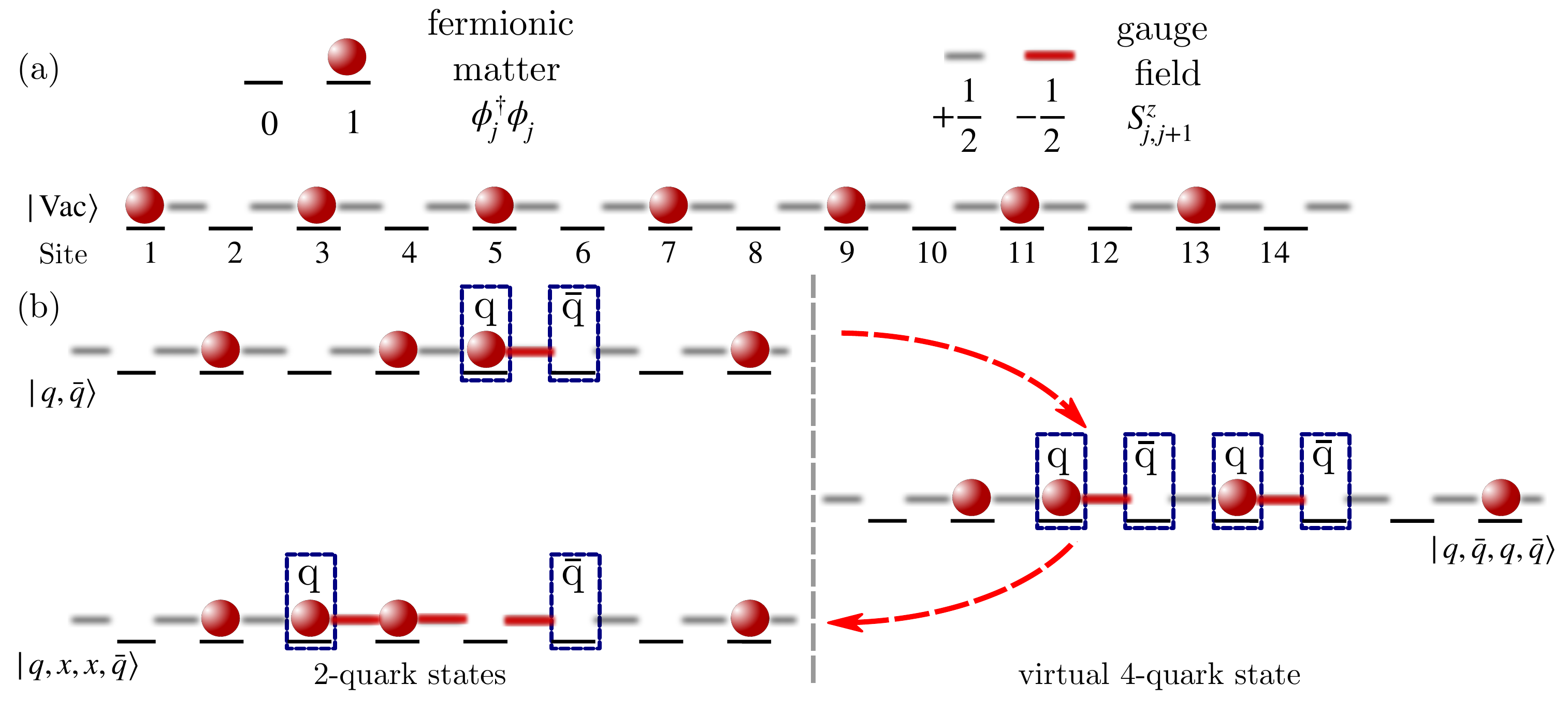}
   \caption{\textbf{Low-energy spectrum of QLM in gauge sector $g_{j}=0$.} (a) We show the vacuum configuration $\ket{\mathrm{Vac}}$ of the $\mathrm{U}(1)$ quantum link model with staggered fermions. Fermionic matter associated to creators $\phi_{j}^{\dagger}$ is located on odd sites and the gauge field takes values of $\langle S^{z}_{j, j+1}\rangle = \frac{1}{2}$ on the links. (b) Application of the gauge-matter coupling of $H_{\mathrm{QLM}}$ allows for creation of quark-antiquark pairs as emphasized in configuration $\ket{q, \bar{q}}$. Dynamical processes of the fermionic excitations ($q$ and $\bar{q}$) can appear via virtual pair-creation processes linking a four-quark configuration $\ket{q, \bar{q}, q, \bar{q}}$ to a two-quark state with larger distance of fermionic excitations $\ket{q, x, x, \bar{q}}$.
    }
    \label{fig:SketchConfigurationsQLM}
\end{figure}

In $(1+1)$-dimensional lattice gauge theory Gauss law imposed by local gauge symmetry on each site completely determines the associated configuration of the gauge field after choosing its values at the boundaries and the desired gauge sector. The latter is defined via the generators of the gauge symmetry 
\begin{equation}
\label{eq:Gauslaw}
    G_{j} = S_{j, j+1}^{z} - S_{j-1, j}^{z} - \phi_{j}^{\dagger} \phi_{j}^{\phantom{\dagger}} + \frac{1- (-1)^{j}}{2}. 
\end{equation}
The set of $\{G_{j}\}_{j}$ commutes with the Hamiltonian, i.e. $[H,G_{j}]=0\;\forall j$, and impose a splitting of the Hilbert space in gauge sectors characterized by $G_{j}\ket{\psi}=g_{j}\ket{\psi}$ and labeled by the eigenvalues $g_{j}$.
Within the zero charge sector $g_j=0$, the vacuum state for the matter is associated with a ferromagnet in the gauge spin degrees of freedom, where $S^z_{j,j+1}$ is aligned in the same direction. For $h_{\parallel}>0$, the ground state has positive magnetization $\langle S^z_{j,j+1}\rangle=+1/2$.
If one now introduces matter particles in the state, the Gauss law forces the gauge spin to change sign each time a quark (or antiquark) is crossed. Hence, quark-antiquark pairs will be separated by gauge spins pointing in the wrong direction: a linear potential proportional to $h_{\parallel}$ is thus established between quark-antiquark pairs, akin to what happens between fermions in the Ising spin chain.
We have seen that in our (1+1)-dimensional QLM fermionic excitations and the configuration of the gauge field are closely related with each other via Gauss law. Therefore, we can simply integrate out the fermionic content of our theory within a given gauge sector and end up in a pure $(1+1)$-dimensional gauge theory.
To better highlight the physics of confinement, we wish to address the effective Hamiltonian in the large mass limit, with the help of \fig{fig:SketchConfigurationsQLM}.

For this we define the rescaled QLM Hamiltonian as
\begin{equation}
\label{eq:rescaledQLM}
    \tilde{H}_{QLM} = \sum_{j=1}^{L} (-1)^{j} \phi_{j}^{\dagger} \phi_{j}^{\phantom{\dagger}} - \frac{4 h_{\parallel}}{m} \sum_{j=1}^{L-1} S_{j, j+1}^{z} -\frac{2\omega}{m} \sum_{j=1}^{L-1}(\phi_{j}^{\dagger} S_{j, j+1}^{+}\phi_{j+1}^{\phantom{\dagger}} + \mathrm{h.c.})  \equiv \tilde{H}^{0} + \tilde{H}^{1}. 
\end{equation}

Whereas Hamiltonian \eq{eq:rescaledQLM} itself does not contain a dynamical term for a single quark, dynamics naturally appears in second order perturbation theory via an intermediate virtual pair-creation process $\mathcal{O}(\frac{\omega}{m})$. In this picture we can go from a configuration $\ket{q, \bar{q}}$ of the particle-antiparticle pair to a configuration with larger distance between the composite particles $\ket{q, x, x, \bar{q}}$ via virtual transition to a four-quark configuration $\ket{q, \bar{q}, q, \bar{q}}$, as emphasized in \figc{fig:SketchConfigurationsQLM}{b}. Notice that we only find a two-site hopping process finally resulting from the staggered nature of the vacuum state.
To write the final effective Hamiltonian in a more compact form, it is useful to introduce the following mapping \cite{Surace2020} between the gauge degrees of freedom and new spin variables
\be
\label{eq:Mapping-QLM-Spins}
     \sigma_{j}^{z} \Longleftrightarrow (-1)^{j} 2 S_{j, j+1}^{z}\, ,\hspace{2pc}  \sigma_{j}^{x} \Longleftrightarrow (\phi_{j-1}^{\dagger} S_{j-1, j}^{+}\phi_{j}^{\phantom{\dagger}} + \mathrm{h.c.}) \,.
\ee
In this notation, the effective Hamiltonian reads
\begin{equation}
\label{eq:eff-spinmodel}
    H^{\mathrm{eff}} = \sum_{j=1}^{L-2} \sigma_{j}^{z}\sigma_{j+1}^{z} - h_{\perp}^{\mathrm{eff}}\sum_{j=1}^{L-2}\sigma_{j}^{x}\sigma_{j+1}^{x} - \frac{2 h_{\parallel}}{m} \sum_{j=1}^{L-1} (-1)^{j} \sigma_{j}^{z}.
\end{equation}
Where the effective transverse field of \eq{eq:eff-spinmodel} is determined as
\be
    -h_{\perp}^{\mathrm{eff}} =  \sum_{\ket{\nu}} \frac{\bra{q_{j-2}, x_{j-1}, x_{j}, \bar{q}_{j+1}}\tilde{H}^{1}\ket{\nu}\bra{\nu} \tilde{H}^{1}\ket{q_{j}, \bar{q}_{j+1}}}{E^{0}_{\ket{q_{j}, \bar{q}_{j+1}}} -  E^{0}_{\ket{\nu}}} 
     = -\frac{4 \omega^{2}}{m} \Big[ 1 + \frac{4 h_{\parallel}}{m^3} \Big].
\ee
Above, we used that the unperturbed energy difference between both states is, moreover, given by $E^{0}_{\ket{q, \bar{q}}} -  E^{0}_{\ket{q_{j-2}, \bar{q}_{j-1}, q_{j}, \bar{q}_{j+1}}} = -\big[m +\frac{4 h_{\parallel}}{m}\big]$.
%

This precisely resembles an antiferromagnetic version of our spin toy model. The energy cost of a ferromagnetic domain wall thereby directly relates to the mass of a quark excitation in the rescaled QLM of \eq{eq:rescaledQLM} and the confining coupling of the gauge field translates to interaction with an external staggered longitudinal field $(-1)^{j} 2 h_{\parallel}/m$.
Having obtained the effective spin Hamiltonian of \eq{eq:eff-spinmodel} we can proceed with quantizing the energy levels of meson-like ($q, \bar{q}$) excitations in similar manner as we did for the Ising chain in previous sections. In the limit of large $m$, where the Ising energy scale dominates contributions of the remaining terms we can once more restrict our considerations to the two particle sector and find a Hamiltonian
\be
\label{eq:2ptHamiltonian-LGT}
H^{\mathrm{AFM}}_{\mathrm{2pt}} = \int\meas{k_{1}}\meas{k_{2}}[\tilde{\epsilon}(k_{1}) + \tilde{\epsilon}(k_{2})] \ket{k_{1}, k_{2}}\bra{k_{1}, k_{2}} + \sum_{j_{1}, j_{2}} \frac{8h_{\parallel}}{m} \vert j_{1} -j_{2}\vert \ket{j_{1}, j_{2}}\bra{j_{1}, j_{2}}
\ee
similar to \eq{eq_2pt_H} for the two-particle problem in the initial Ising model, where we choose to write down the confining part of \eq{eq:2ptHamiltonian-LGT} in its real space representation. The kinetic part diagonal in the basis of momentum states $\ket{k_{1}, k_{2}}$, however, now contains the dispersion law of the antiferromagnetic Ising chain ($\tilde{\epsilon}(k)\approx 2 - 2h_{\perp}^{\mathrm{eff}}\cos(2k) + \dots$). With this we can proceed in analogous fashion to section~(\ref{SM_sec_conf_is}) and obtain the energy bands for the mesonic exciations using suited variations of \eqsto{eq:H_rel}{eq:QuantizationDisperionsBands}. 
\bigskip

\paragraph{Implementation in Rydberg atoms arrays.---} As pointed out in Ref. \cite{Surace2020}, the Quantum Link Model finds a natural implementation in Rydberg atoms arrays. To connect the two setups, one relies once again on the mapping of \eq{eq:Mapping-QLM-Spins}, but this time we will not use perturbation theory and directly map the Hamiltonian of \eq{eq:QLM} to the Fendley-Sengupta-Sachdev Hamiltonian \cite{Fendley2004} with an additional staggered field
\be
H_\text{Ryd}=\mathcal{P}\left[\sum_j -\omega\sigma_j^x-\frac{m}{2} \sigma_j^z+ h_{\parallel} (-1)^j \sigma_j^z \right]\mathcal{P}\, .
\ee
Where the ground state and Rydberg excited state of each atom are respectively denoted with the spin down and up in the $z-$basis.
Above, the projector $\mathcal{P}$ enforces the Rydberg blockade, namely two nearby atoms cannot be simultaneously excited due to energetic constraints, which is a direct consequence of the Gauss law of the original Quantum Link Model.
As we discussed, in the large mass limit the vacuum of the gauge theory in the zero charge sector consists in a $z-$ferromagnet in the gauge field. Therefore, through the mapping of \eq{eq:Mapping-QLM-Spins}, this state is mapped to a Rydberg configuration where atoms are excited on alternating bonds, the weak staggered coupling $\propto h_{\parallel}$ breaks the $\mathbb{Z}_2$ symmetry favoring one staggered configuration in place of the companion related by a one-site shift.
Quark-antiquark pairs are in correspondence with defects in this staggered configuration: by acting on the vacuum with a weak rotation along the $x-$direction $R= e^{i\theta\sum_j\sigma_j^x }$ with $\theta\ll 1$ it is possible to create a homogeneous gas of quark-antiquark pairs with zero total momentum, akin of the initial state produced in the Ising chain.
Such an excited state will then relax through the multistage thermalization dynamics we exhaustively discussed in the Ising chain.

\section{Exact diagonalization in the few kinks subspace}
\label{SM_sec_ED}

Building on the stability of the fermions for exponentially long times, one can project the Hamiltonian of \eq{eq_2pt_H} within the few-fermions sector. In this way, by exact numerical integration of the few-fermion wavefunction, we can access very long timescales and explore prethermalization.

While this strategy can be applied for arbitrary values of the transverse field by considering \eq{eq_2pt_H} (generalized to many fermions), here, we focus on the small transverse field limit where the Hamiltonian is further simplified.
In this regime, the kinetic part reduces to nearest-neighbor hopping and the fermions are equivalently describing domain walls.
Hence, let $\Psi(j_1,j_2,...,j_{2n-1},j_{2n})$ be the wave function labeling the state with domain walls between the lattice sites $j_{i}-1$ and $j_i$ and, without loss of generality, we consider the ordering $j_1<j_2<...<j_{2n}$, and periodic boundary conditions are assumed.
Furthermore, we consider the false vacuum to be  between the kinks $j_{2i-1}$ and $j_{2i}$.
On this wave funtion, the Hamiltonian acts as
\be\label{eq_H_multikink}
[\hat{H}_\text{Kinks}\Psi](j_1,...,j_{2n})=\sum_{i=1}^{2n}-\hpe[\Psi(j_1,...,j_i+1,...,j_{2n}) +\Psi(j_1,...,j_i-1,...,j_{2n}) ]+\sum_{i=1}^{n}\chi |j_{2i-1}-j_{2i}|\Psi(j_1,...,j_{2n})\, .
\ee
Above, we neglect an overall unimportant constant and the hopping term should respect the hard core constraint $j_1<j_2<...<j_{2n}$.
Since we are mostly interested in the scattering among mesons, we consider a translational invariant scenario: this allows us to further enhance the performance of the approach by removing a degree of freedom.
\begin{figure}
    \centering
    \includegraphics[width=\textwidth]{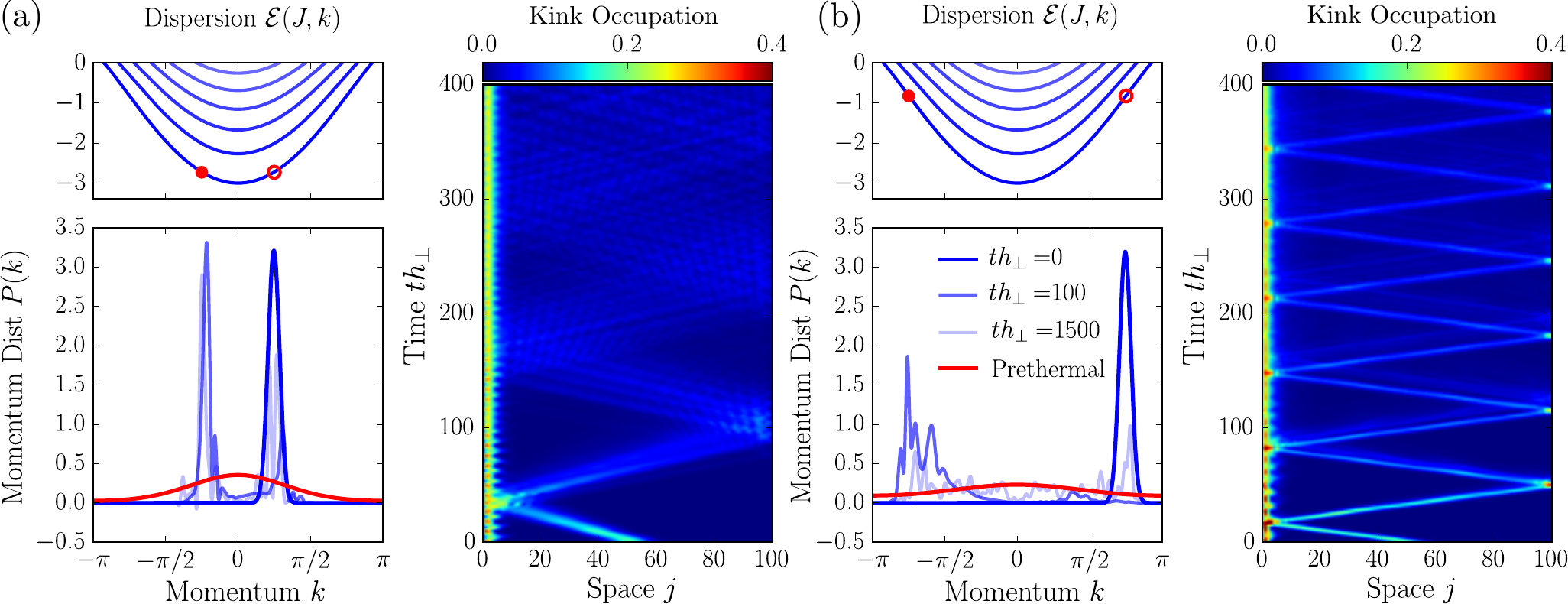}
    \caption{\textbf{Two-meson dynamics.}
    Analogously to the three-meson scenario considered in the main text, we illustrate the results for dynamics of two mesons in a system of $L=100$ sites and confinement field $\hpa/\hpe=0.1$. We consider two distinct initial states, where the mesons are initialized with energies below (a) or above (b) the second band of the single meson spectrum $\mathcal{E}(J, k)$.
    The momentum distribution $P(k)$ of the meson with positive initial momentum is evaluated (empty dot in the plot of dispersion bands).  (a) A double-peak structure centered at the initial momenta of the mesons $k_{0}=\pm\pi/4$ survives even at late times. This characteristics disappears in (b) for initial meson momenta of $(k_{0}=\pm3\pi/4)$, where after a time transient various momenta are be occupied. Crucially, for both initial states, (a) and (b), the system does not attain the prethermal state (red line) because scattering events involve only two mesons.
    This fact clearly distinguishes the scenario of  two mesons from the case of three mesons discussed in the main text.}
    \label{fig:S_1}
\end{figure}
For the sake of simplicity, we consider the case of global zero momentum, but the same method can be applied to the general case.
It is convenient to use the position of the first domain wall as a reference coordinate and introduce new variables $s_i=j_{i+1}-j_1$. We denote with $\Phi(s_1,...,s_{2n-1})$ the wavefunction in the relative coordinates. In this case, the dynamics is
\begin{multline}
[\hat{H}_{\text{Kinks},\text{zero momentum}}\Phi](s_1,...,s_{2n-1})=-\hpe[\Phi(s_1+1,...,s_{2n-1}+1)-\Phi(s_1-1,...,s_{2n-1}-1)]+\\
\sum_{i=1}^{2n-1}-\hpe[\Phi(s_1,...,s_i+1,...,s_{2n-1}) +\Phi(s_1,...,s_i-1,...,s_{2n-1}) ]+\sum_{i=0}^{n-1}\chi |s_{2i+1}-s_{2i}|\Phi(s_1,...,s_{2n-1})\, .
\end{multline}
Above, the first term accounts for the hopping of the first domain wall, which equivalently shift of one site all the relative distances $s_i$.
From the knowledge of the wavefunction $\Phi$, several observables of interest can be computed.
First, the total magnetization is directly connected to the length of the mesons, since the part of the chain enclosed within one meson lays in the false vacuum
\be
\sum_{j=1}^L\langle S^z_j\rangle=L/2 -\sum_{\{s_i\}}\left[s_1+\sum_{i=1}^{n-1}|s_{2i+1}-s_{2i}|\right]|\Phi(s_1,...,s_{2n-1})|^2\, .
\ee
However, much more information is contained in the probability distribution of the meson length
\be
\mathcal{P}_{\text{Length}}(\ell)=\frac{1}{N_{\text{mes}}}\sum_{\{s_i\}}\left[\delta(\ell-s_1)+\sum_{i=1}^{n-1}\delta(\ell-|s_{2i+1}-s_{2i}|)\right]|\Phi(s_1,...,s_{2n-1})|^2\, ,
\ee
where $\delta$ is a Kronecker delta distribution.
Nonetheless, our primary tool to assess prethermalization is the momentum distribution of the mesons. In this case, particular care should be taken when passing from the original coordinates to the relative ones.
Let us consider the density matrix in the momentum space defined as
\be\label{eq_S33}
\rho(k_1,k_2,...|q_1,q_2,...)=\sum_{\{j_i\},\{j_i'\}}e^{i\sum_i (k_i j_i-q_i j_i')}\Psi(j_1,j_2,...)\Psi^*(j_1',j_2',...)\, .
\ee
Then, we wish to target the momentum distribution of the first meson $P(k)$, defined as
\be
P(k)=\sum_{\{k_i\}}\delta(k_1+k_2-k)\rho(k_1,k_2...|k_1,k_2,...)=\int \frac{\dd\omega}{2\pi}\sum_{\{k_i\}}e^{i\omega(k_1+k_2-k)}\rho(k_1,k_2...|k_1,k_2,...)\,.
\ee
The integral representation of the Dirac $\delta$ distribution is particularly convenient for carrying out the straightforward but lengthy calculations. By plugging the definition of \eq{eq_S33} in the above equation and, within the total zero momentum sector, passing to the relative coordinates one finally obtains
\be
P(k)=\sum_{\{s_i\}_{i=1}^{2n}}e^{iks_1}\Phi(s_2-s_1,s_3-s_1,...,s_{2n}-s_1)\Phi^*(s_2-s_1,s_3,...,s_{2n})\, .
\ee
\begin{figure}[t!]
    \centering
    \includegraphics[width=\textwidth]{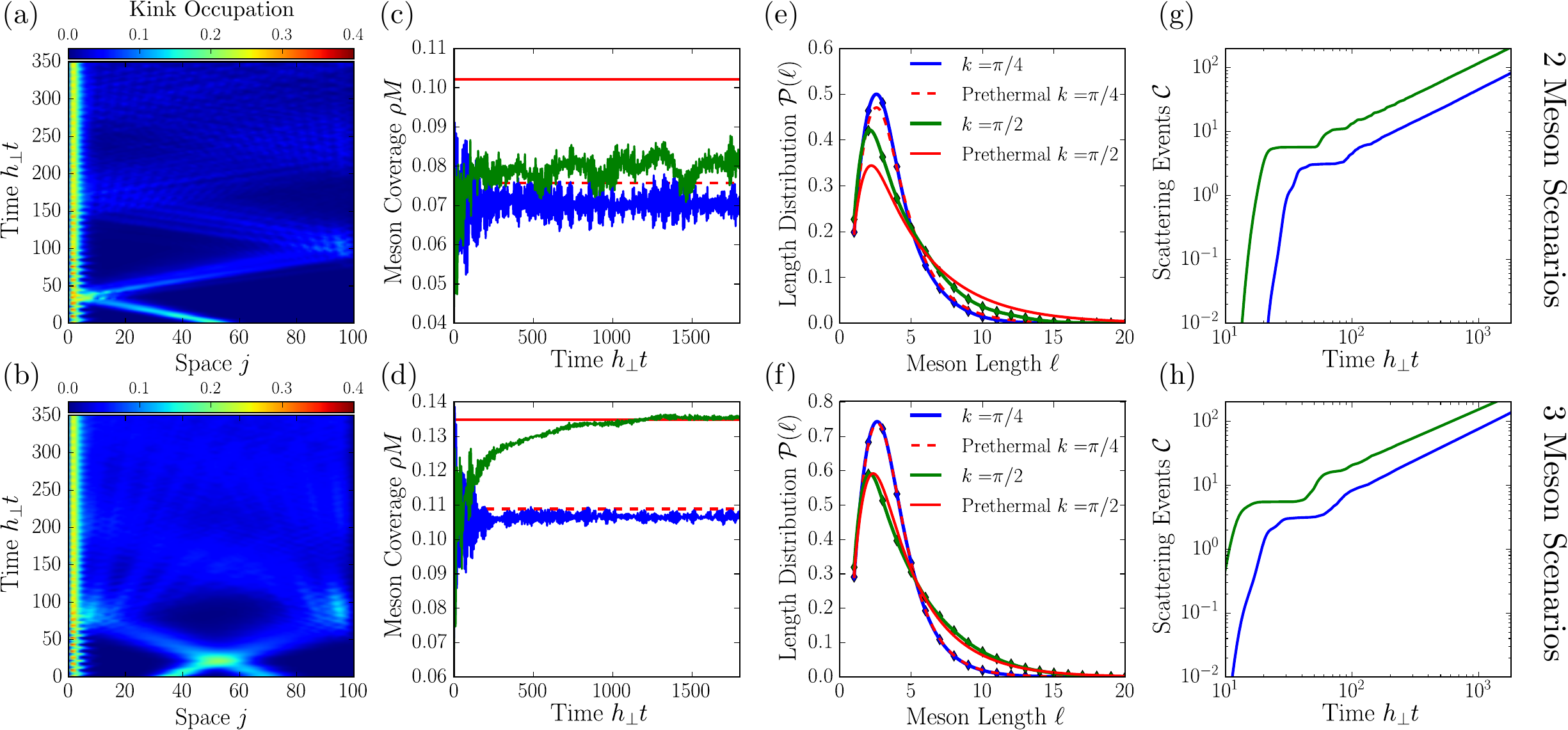}
    \caption{\textbf{Comparison of two and three meson scenario.} 
    Dynamics in a system of $L=100$ sites containing (a) two mesons with momenta $k_{0}=\pm\pi/2$ and (b) three mesons with momenta $k_{0}=0,\pm\pi/2$, respectively. The underlying confinement field is $\hpa/\hpe=0.1$. The meson coverage $\rho M$ reveals a fundamental difference between systems of (c) two and (d) three mesons. Whereas an ensemble of three mesons relaxes to a prethermal configuration (red dashed and solid lines) provided the initial energy per meson is above the second band of the single meson spectrum, we solely find relaxation to a non-thermal state for the two meson system. (e) - (f) This observation is supported by studies of the length distribution of mesons $\mathcal{P}(\ell)$. (g) - (h) Even though the meson densities are different in the two cases, the number of scattering events is similar. Therefore, an absence of thermalization in the two-meson scenario, due to a reduced number scattering events can be ruled out.}
    \label{fig:S_2}
\end{figure}

In the main text, we use the momentum distribution to analyze relaxation and prethermalization of initial wavepacket configurations.
To this end, we initialize states with two ($n=2$) or three mesons ($n=3$) in the form of gaussian wavepackets with tunable initial momenta. The functional form of the initial meson state reads
\be
\label{eq:initial-state}
\Phi(s_1, ..., s_{2n}) = \prod_{i=0}^{n-1} \left\{ \phi_{K_{i}}(s_{2i+1}-s_{2i}) e^{i K_{j}(\frac{s_{2i+1}+s_{2i}}{2})}
W_{\sigma, \bar{X}_i}\left(s_{2i}, s_{2i+1}; s_{2i+2}, s_{2i+3} \right)\right\} ,
\ee
where above $W_{\sigma,X}$ is a gaussian wavepacket for the relative distance between to consecutive mesons
\be
W_{\sigma,X}(s_{2i},s_{2i + 1};s_{2i + 2},s_{2i+3}) \propto \exp\Bigg[-\frac{\big((s_{2i} + s_{2i + 1}) - (s_{2i + 2} + s_{2i+3}) - \bar{X}\big)^2}{\sigma^2} \Bigg]
\ee
and the wavefunction $\phi_K(s)$ is the mesonic wavefunction of the lowest dispersion band obtained by numerically diagonalizing \eq{eq_H_multikink} in the two-fermion sector with total momentum $K$. Furthermore, we insert a cutoff $\phi_K(|s|<\lambda_c)=0$ (and similarly in the gaussian wavepackets) to ensure the initial state is correctly ordered. We checked our results to be cutoff-independent.

Further results of the scattering dynamics are provided in supplementary figures \ref{fig:S_1} and \ref{fig:S_2}.

\begin{figure}[t!]
    \centering
    \includegraphics[width=0.9\textwidth]{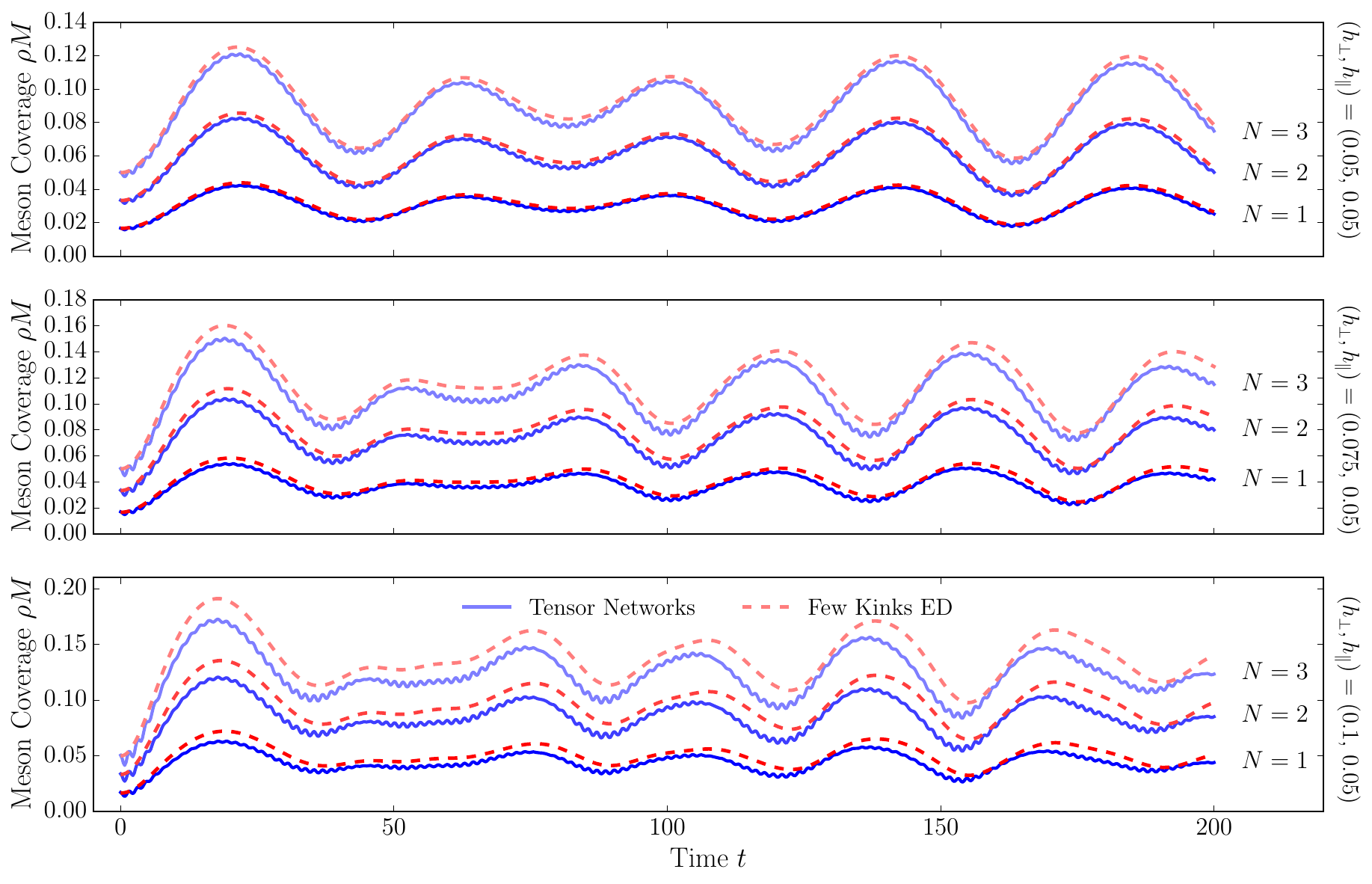}
    \caption{\textbf{Comparison of few-kink subspace evolution with tensor network predictions.} We benchmark the evolution results for a $N$ meson initial state $\ket{\psi_{i}^{(N)}}$ obtained using exact diagonalization in the few kink subspace against numerically exact results from tensor network evolution. We consider values for the transverse field of $h_{\perp} \in \{0.05, 0.075, 0.1\}$ and fixed longitudinal field $h_{\parallel}=0.05$ while numbers of $N\in \{1, 2, 3\}$ mesons in the chain are tested. We find very good agreement between results in systems of $L=60$ sites and bond dimensions up to $\chi=256$ for small values of $h_{\perp}=0.05$. Considering larger values of the transverse field $h_{\perp}=0.1$ we find qualitative features of the tensor network evolution to be reproduced by exact diagonalization results, while still differing by a offset expected from dressing of the fermionic domain walls at larger values of $h_{\perp}$.}
    \label{fig:Benchmark_ED_vs_TN}
\end{figure}

\subsection{Benchmarking exact diagonalization in the few kinks subspace with tensor networks}
\label{TN_vs_FKED}

After discussing the basics of both exact diagonalization in the few kinks subspace and tensor network simulations it proofs useful to compare both methods with each other to benchmark the accuracy of exact diagonalization results. Since we refer to results of few kinks ansatz in the main text for situations considering up to 3 mesons in the system we would like to test evolution containing $N\in\{1, 2,3\}$ mesons. As one of the main assumptions leading to the formulation of the few kinks ansatz, moreover, was the sharp nature of fermionic domain wall excitations, the accuracy will crucially depend on the chosen transverse field $h_{\perp}$ dressing the sharp fermionic kinks. For this reason we also provide simulation results for different choices of $h_{\perp}\in\{0.05, 0.075, 0.1\}$. The value of the longitudinal field component $h_{\parallel}$ will thereby be kept constant with at a rather small value of $h_{\parallel}=0.05$ to guarantee conservation of the meson number. As a suitable initial state for our comparison we implement a homogeneous superposition of isolated single spin flips applied to the ground state $\ket{0}$ of the system for given values of $(h_{\parallel}, h_{\perp})$. This can be formally written as
\begin{equation}
\label{eq:InitialState}
\ket{\psi_{i}^{(N)}} = \sum_{\{j_{1}, j_{2}, ..., j_{N}\}} \sigma_{j_{1}}^{x} \sigma_{j_{2}}^{x} \dots \sigma_{j_{N}}^{x} \ket{0}.
\end{equation}
$\{j_{1}, j_{2}, ..., j_{N}\}$ thereby describes a ordered tuple of integers taking values between 2 and $L-$1 with differences of neighboring integers $j_{n}-j_{n-1}>1$ for $2\le n \le N$ in order to realize isolated spin flips not located at the boundary. For small values of the transverse field $h_{\perp}$ this state has large overlap with a $N$ meson configuration. Each meson thereby has the smallest possible extent of a single lattice spacing. The results for the evolution using few kinks exact diagonalization and tensor network methods in a system with open boundary conditions are shown in \fig{fig:Benchmark_ED_vs_TN}. For comparison we investigate the meson coverage $\rho M$, which is following the convention of the main text closely related to the magnetization in the system. We find very good agreement between both methods for all tested values of $N\in\{1, 2, 3\}$ and small values of $h_{\perp}=0.05$. We, moreover, find qualitative features of the evolution like the oscillation frequency of $\rho M$ preserved also for larger values of $h_{\perp}=0.1$. We can, moreover, identify an additional offset between results of tensor networks respectively to predictions of exact diagonalization though. This deviation is in agreement with our expectations of decreasing accuracy for larger values of $h_{\perp}$. This indicates that simulations using exact diagonalization in the few kinks subspace can be used to characterize the physics of a system of few mesons in the limit of small transverse field $h_{\perp}$. 

\twocolumngrid
\bibliography{biblio}


\end{document}